\documentclass[paper,prd,reprint,preprintnumbers,nofootinbib,notitlepage,showpacs,showkeys]{revtex4-1}
\usepackage{latexsym,graphicx,amssymb,amsmath,mathrsfs} 

\DeclareSymbolFont{bbold}{U}{bbold}{m}{n}
\DeclareSymbolFontAlphabet{\mathbbold}{bbold}

\newcommand{\be}{\begin{equation}}      
\newcommand{\ee}{\end{equation}}      
\newcommand{\bea}{\begin{eqnarray}}      
\newcommand{\eea}{\end{eqnarray}}    
     
\newcommand{\rt}[1]{{}}      
  
\newcommand{\ifff}{\,\textrm{if}\,}  
\newcommand{\elsee}{\,\textrm{else}\,}  
\newcommand{\Tr}{\,\textrm{Tr}\,}
\newcommand{\STr}{\,\textrm{STr}\,}
\newcommand{\diag}{\,\textrm{diag}\,} 
\newcommand{\pion}{\,\textrm{pion index}\,} 
\newcommand{\kaon}{\,\textrm{kaon index}\,}
\newcommand{\ife}{\,\textrm{if}\,}
\newcommand{\orr}{\,\textrm{or}\,} 
\newcommand{\els}{\,\textrm{else}\,}

\makeatletter
\renewcommand\appendix{\par
\setcounter{section}{0}%   
\setcounter{subsection}{0}% 
\gdef\thesection{\appendixname\space\@Alph\c@section}}

\long\def\unmarkedfootnote#1{{\long\def\@makefntext##1{##1}\footnotetext{#1}}}
\makeatother

\begin{document} 
%{\allowdisplaybreaks

\title{Fluctuation induced first order phase transition in $U(n)\times U(n)$ models \\ using chiral invariant expansion of functional renormalization group flows} 

\author{G. Fej\H{o}s}
\email{fejos@riken.jp}
\affiliation{Theoretical Research Division, Nishina Center, RIKEN, Wako 351-0198, Japan}
\preprint{RIKEN-QHP-161}

\begin{abstract}
{Phase transition in $U(n)\times U(n)$ models is investigated for arbitrary flavor number $n$. We present a nonperturbative, 3+1 dimensional finite temperature treatment of obtaining the effective potential, based on a chiral invariant expansion of the functional renormalization group flows. The obtained tower of equations is similar but not identical to that of the Dyson-Schwinger hierarchy and has to be truncated for practical purposes. We investigate the finite temperature behavior of the system in an expansive set of the parameter space for $n=2,3,4$ and also perform a large-$n$ analysis. Our method is capable of recovering the one-loop $\beta$ functions of the coupling constants of the $\epsilon$ expansion; furthermore, it shows direct evidence that regardless of the actual flavor number, within our approximation, the system undergoes a fluctuation induced first order phase transition.}
\end{abstract}

\pacs{11.30.Qc, 11.30.Rd}
\keywords{Chiral symmetry breaking, functional renormalization group, first order phase transition}  
\maketitle

\section{Introduction}

Spontaneous breaking of chiral symmetry of quantum chromodynamics (QCD) can be studied efficiently via effective scalar theories. The $U(n)\times U(n)$ family of meson models, based on (approximate) $n$-flavor chiral symmetry is a popular starting point of investigating low-energy strong interaction \cite{gellmann60,chan73,lenaghan00,roder03,herpay06,kovacs07,kovacs08,schaefer09,jakovac10,mitter13}, giving account of the spontaneous breaking $U_L(n)\times U_R(n)\longrightarrow U_V(n)$. Depending on the actual energy scale, flavor numbers $n=2,3$ have phenomenological relevance, with an unlikely extension to $n=4$. Temperatures around the scale of QCD, $\Lambda_{QCD}$, the symmetry restores, but the details of this transition are far from being understood theoretically, especially in high density regions. For instance, its properties strongly depend on the strength of the anomalous breaking of the $U_A(1)$ subgroup and its temperature dependence \cite{fukushima10}. One might also be interested in the quark mass sensitivity of the properties of the transition; see discussions on the Columbia plot \cite{fukushima10}.

With the use of the $\epsilon$ expansion, in the pure $U(n)\times U(n)$ model [i.e. without explicit symmetry breaking terms and vanishing $U_A(1)$ anomaly] Pisarski and Wilczek calculated the $\beta$ functions and showed that there was no infrared stable fixed point for any flavor number $n\geq 2$ \cite{pisarski84}. This serves as indirect evidence of a phase transition, if it exists, being fluctuation induced, and of first order. This conjecture has been confirmed by various nonperturbative studies for the $n=2$ case, especially with the use of the functional renormalization group (FRG) \cite{berges02,fukushima10b}; however, investigations of this issue for flavor numbers $n>2$ are virtually missing from the literature.

Even though correctness of the $\epsilon$ expansion is widely accepted, its conclusions have to be treated cautiously. There are counterexamples in which the method breaks down and its prediction of a fluctuation induced first order transition fails; see the example of superconductivity \cite{kleinert06}. Concerning the current model at $n=2$, there are also signs of having a stable infrared fixed point after all; see investigations based on perturbative \cite{pelissetto13} and conformal bootstrap methods \cite{nakayama14}. We also note that recently this issue has been under investigation concerning two-flavor QCD itself \cite{aoki12,aoki14}.

In this paper we are looking for a direct evidence of a fluctuation induced first order transition in the $U(n)\times U(n)$ models, irrespective of the flavor number characterizing the group structure. We wish to confirm the results of the $\epsilon$ expansion, and gain a deeper understanding of the mechanism from a nonperturbative point of view. We are going to work in the functional renormalization group formalism and solve the resulting flow equations nonperturbatively. By nonpertubativity we mean that fully functional solutions will be considered, not only the flow of Taylor coefficients as individual coupling constants. We would like to emphasize the importance of this procedure, since as argued in \cite{fukushima10b,nakano09}, Taylor-expanding the flow equations up to a few orders around the minimum of the effective potential might give qualitatively correct insight of its evolution, but in the case of a first order transition, it cannot reproduce the full solution in a quantitative way.

Unlike earlier studies carried out in 3 dimensions with formally temperature dependent parameters \cite{berges02,fukushima10b}, our treatment is formulated in $3+1$ dimensions, with the explicit presence of the temperature. This is motivated by the fact that strictly speaking, in the compactified (Euclidean) time direction of the path integral, long range fluctuations are unable to propagate only when the system is close to criticality; therefore the dimensionally reduced theory can only be relevant if a second order transition is present. If criticality is absent in the model, in order to develop a reliable method for phenomenological applications, explicit introduction of the temperature is necessary. As implied, the paper is also trying to contribute as a methodological study towards phenomenology, but leaves the completion of this task for a further publication.

The paper is organized as follows. In Sec. II, we define the model and the corresponding notations. We derive the flow equation of the scale dependent effective potential $V_k$ and present a systematically improvable approximation scheme based on a chiral invariant expansion. With the use of this formulation, we reproduce the one-loop results of the $\epsilon$ expansion for arbitrary flavor number $n$ using the FRG method. In Sec. III, we solve the resulting equations numerically, present the results and map an expansive set of the parameter space of the model for flavor numbers that might be relevant for phenomenology (i.e., $n=2,3,4$). We also investigate briefly the order of the transition for higher flavor numbers and perform a large-$n$ analysis. Finally, in Sec. IV, the reader can find some remarks and the conclusions.

\section{Flow equations and approximation schemes}

\subsection{Basics}

The $U(n)\times U(n)$ symmetric scalar model is a theory of a matrix field $M\in {\cal G}(U(n))$
\bea
M=(s^a+i\pi^a)T^a, \qquad (a=0...,n^2-1),
\eea
where $s^a$, $\pi^a$ coefficients are the scalar and pseudoscalar fields, respectively, with $T^a$ being the group generators following the convention $\Tr (T^a T^b)=\delta^{ab}/2$. The Lagrangian is
\bea
\label{Eq:Lag}
{\cal L}&=&\partial_\mu M \partial^{\mu} M^\dagger - \mu^2 \Tr (MM^\dagger) \nonumber\\
&-&\frac{g_1}{n^2}[\Tr (MM^\dagger)]^2- \frac{g_2}{n}\Tr (MM^\dagger M M^\dagger),
\label{Eq:lag}
\eea
where we choose $\mu^2<0$. Expression (\ref{Eq:Lag}) is symmetric under $U(n)\times U(n)$ rotations 
\bea
M \longrightarrow R M L^\dagger \quad \quad (L,R \in U(n)),
\eea
or in terms of pure vector and axial-vector transformations,
\bea
M \longrightarrow V^\dagger M V, \quad M\longrightarrow A^\dagger M A^\dagger, \quad [A,V \in U(n)],
\eea
where the following relations hold between the transformation parameters: $\theta^a_{V,A}=(\theta^a_L\pm \theta^a_R)/2$. The classical potential has to be bounded from below, which restricts the coupling space as $g_1+g_2>0$ and $g_1+ng_2>0$ \cite{fejos13}.  It is known that in terms of symmetry breaking the model is divided into two parts: if $g_2>0$, the symmetry breaks as $U(n)\times U(n) \longrightarrow U(n)$, while $g_2<0$ leads to $U(n)\times U(n) \longrightarrow U(n-1)\times U(n-1)$ \cite{paterson81}. Even though in this paper the pure $U(n)\times U(n)$ model [i.e. without explicit symmetry breaking terms and $U_A(1)$ anomaly] is investigated, our final goal is to see the applicability of the extensions of the model to low-energy QCD; therefore we restrict ourselves to $g_2>0$, which obeys the observed (approximate) chiral symmetry breaking of nature.

As already stressed in the Introduction, in the present work we investigate the finite temperature properties of the system (\ref{Eq:lag}) in $3+1$ dimensions, with the use of the functional renormalization group method \cite{wetterich93,pawlowski07}. The main concept of the formalism is to introduce a scale dependent effective action $\Gamma_k$, which obeys the flow equation
\bea
\partial_k \Gamma_k = \frac12 \STr \left[\frac{1}{\Gamma_k^{(2)}+R_k}\partial_k R_k\right],
\label{Eq:flow}
\eea
where the $\STr$ operation has to be taken both in functional and matrix sense; $\Gamma_k^{(2)}$ is the second functional derivative of $\Gamma_k$ with respect to the dynamical fields and $R_k$ is a regulator function. The scale dependent effective action $\Gamma_k$ interpolates between the classical (microscopic) action (at $k=\Lambda$, being an appropriately high UV scale) and the quantum effective action (at $k=0$). Equation (\ref{Eq:flow}) is an exact relation if the regulator function is chosen to be suppressing IR modes below scale $k$, but it has to be approximated for practical purposes. In this study we use the local potential approximation (LPA), which is the leading order contribution of the derivative expansion; therefore $\Gamma_k$ is approximated as an integral of a local function depending on the average fields $\bar{M}$ as
\bea
\Gamma_k[\bar{M}] = \int d^4x \left(\partial_\mu \bar{M}(x)^\dagger \partial^\mu \bar{M}(x) - V_k(x;\bar{M})\right).
\eea
We use Litim's 3D regulator \cite{litim01}:
\bea
\label{Eq:reg}
R_k(p)=(k^2-{\bf{p}}^2)\Theta(k^2-{\bf{p}}^2),
\eea
where we note immediately that this regulator does not lead to any scale separation in the timelike direction (i.e. the frequency space), and therefore $\partial_k R_k$ in the flow equation (\ref{Eq:flow}) serves as a UV cutoff only in spacelike directions. This may raise doubts on the applicability of the derivative expansion; nevertheless, it has been shown that in $O(N)$ symmetric theories LPA works extremely accurately with even this type of regulator \cite{blaizot06,blaizot10}. We expect that it might lead to reasonable results in the current model as well. Similarly as it was done in the case of the $O(N)$ model \cite{blaizot10}, the clarification of this issue would require mapping the regulator dependence, but we leave this question for further studies.

Using (\ref{Eq:reg}) in (\ref{Eq:flow}), in $d$ (spatial) dimensions, at finite temperature $T$, we get the following flow equation for the potential $V_k[\bar{M}]$:
\bea
\partial_k V_k[\bar{M}]=k^{d+1}K_d T\sum_{\omega_m}\sum_i \frac{1}{\omega_m^2+k^2+\mu_i^2(k)},
\label{Eq:flow_Vk}
\eea
where $K_d^{-1}=2^{d-1}\pi^{d/2}\Gamma(d/2)d$. We have to sum over bosonic Matsubara frequencies $\omega_m=2\pi m T$, and $\mu^2_i(k)$ denotes the eigenvalues of mass matrices $\partial^2V_k/\partial s^i\partial s^j$ and $\partial^2V_k/\partial \pi^i \partial \pi^j$.

The $V_k$ local potential cannot depend on arbitrary $\bar{M}$ configurations, it has to reflect the $U(n)\times U(n)$ symmetry, which means that its actual variables are $U(n)\times U(n)$ invariants in a background field $\bar{M}$:
\bea
V_k(\bar{M})\equiv V_k(I_1,I_2...,I_n).
\eea
A possible set of independent chiral invariants has been denoted by $I_i$. We use the following basis:
\bea
I_1&=&\Tr (\bar{M}^\dagger \bar{M}), \quad I_2=\Tr [\bar{M}^\dagger \bar{M}-\Tr(\bar{M}^\dagger \bar{M})/n]^2, \nonumber\\
&&... \quad I_n=\Tr [\bar{M}^\dagger \bar{M}-\Tr(\bar{M}^\dagger \bar{M})/n]^n.
\eea
Before solving (\ref{Eq:flow_Vk}), a particular symmetry breaking pattern has to be chosen as well. One may start with the most general scenario, and build up the condensate $\bar{M}$ as a linear combination of diagonal generators (as $\bar{M}$ can always be thought as a diagonal matrix, since with the application of appropriate vector transformations it is possible to diagonalize it without changing the Lagrangian):
\bea
\bar{M}=\sum_{i=\diag} v_i T^i.
\label{Eq:gen_backg}
\eea
However, it was shown in \cite{fejos13} that even though there are multiple ($2^n$) minima of the effective potential in the diagonal condensate space, they are equivalent and their corresponding directions are connected via axial-vector transformations. We can therefore assume that for searching symmetry breaking minima, without the loss of generality,
\bea
\bar{M} = v_0 T^0,
\label{Eq:v0_backg}
\eea
and investigate the flow of $V_k$ in a one-component condensate space. This seems to be particularly convenient as all invariants but $I_1$ are zero in the background field (\ref{Eq:v0_backg}). Concerning $V_k$, one nevertheless also needs the dependence on $I_{i\geq 2}$, as it is shown in the next subsection.

\subsection{Invariant expansion and flow hierarchy}

Since we are interested in a symmetry breaking pattern where the condensate is proportional to the unit matrix, it is expected to be useful expanding $V_k$ around this configuration. In terms of chiral invariants, this leads to the following expression:
\bea
V_k(I_1,I_2,...I_n) = U_k(I_1) + \sum_{\{\alpha\}} C_k^{(\alpha)}(I_1) \prod_{i=2}^{n}I_i^{\alpha_i},
\label{Eq:Vk_inv}
\eea
where $\alpha$ is a multi-index with $n$ entries, denoting the occurrences of invariants in each term of the expression. For the sake of an example, $\alpha=(0,3,2,1,0,...,0)$ refers to the term $C_k^{(0,3,2,1,0,...0)}(I_1)\cdot I_2^3 I_3^2 I_4$ (the first entry of a given $\alpha$ is always zero, since the expansion is realized as a power series in $I_{i\geq 2}$). Note that $I_1$  is the only nonzero invariant in a background field defined by $v_0$. The reason for keeping terms besides $U_k(I_1)$ is because, as we will see, its flow entangles with the flow of other $C_k^{(\alpha)}$ coefficients. For the evaluation of $V_k$ in the background field (\ref{Eq:v0_backg}), only $U_k(I_1)$ is needed, since $I_{i\geq 2}|_{v_0}\equiv 0$.

The invariant expansion (\ref{Eq:Vk_inv}) is very convenient from a numerical point of view, since instead of calculating a function in an $n$-dimensional grid, we build up flow equations for $1$-dimensional coefficient functions. The most general steps of obtaining these equations are as follows. At first one has to consider the most general symmetry breaking pattern (\ref{Eq:gen_backg}) and calculate the mass matrices and their eigenvalues from the effective potential. After substituting these eigenvalues into the right-hand side of (\ref{Eq:flow_Vk}), it has to be expanded around $\partial_k V_k(I_1,0,0,...)\equiv \partial_k U_k(I_1)$. Since the masses are not chiral invariants, as opposed to the flow equation, the obtained terms (yet in the language of condensates) must be combinable into a $U(n)\times U(n)$ invariant expression \cite{patkos12}, ultimately taking the form of (\ref{Eq:Vk_inv}). This completes the construction; the flow of coefficients $C_k^{(\alpha)}(I_1)$ are now ready to be identified.

Let us see how this works in practice. At first, let us start with the flow of $U_k$. To obtain this, we do not  actually have to consider the most general background field; inserting (\ref{Eq:v0_backg}) into (\ref{Eq:flow_Vk}) is sufficient, since $I_{i\geq 2}|_{v_0}=0$. Using that $I_1|_{v_0}=v_0^2/2$, we get the following spectrum:
\begin{subequations}
\label{Eq:masses}
\bea
\mu_{a_0}^2(k)&=&U_k'(I_1)+\frac{4I_1}{n}C_k^{(0,1,0,0...)}(I_1), \hspace{1.5cm} \\
\mu_{\sigma}^2(k)&=&2I_1U_k''(I_1)+U_k'(I_1), \hspace{3.1cm} \\
\mu_{\pi}^2(k)&=&U_k'(I_1), \hspace{5cm} 
\eea
\end{subequations}
where the multiplicities are $n^2-1$ $(a_0)$, $1$ $(\sigma)$, and $n^2$ $(\pi)$, respectively. The pseudoscalars ($\pi$) receive the same mass, whereas the scalars ($a_0$, $\sigma$) split. Inserting (\ref{Eq:masses}) into (\ref{Eq:flow_Vk}), we obtain
\bea
\label{Eq:flow_Uk}
\!\!\!\!\!\partial_k U_k(I_1)&&=k^{d+1}K_d T \times \nonumber\\
&&\sum_{\omega_m}  \left(\frac{n^2}{\omega_m^2+E_\pi^2}+\frac{n^2-1}{\omega_m^2+E_{a_0}^2}+\frac{1}{\omega_m^2+E_\sigma^2}\right),
\eea
where $E_i^2=k^2+\mu_i^2(k)$ $(i=\pi,a_0,\sigma)$. Note that the mass of $a_0$ contains the coefficient $C_k^{(0,1,0,0...)}(I_1)$; therefore evaluating the functional flow equation (\ref{Eq:flow}) in the background field $v_0$ does not give a closed equation for $U_k(I_1)$, as one also needs to know the flow of $C_k^{(0,1,0,0...)}(I_1)$. One can easily see that a flow of a given coefficient in the invariant expansion (\ref{Eq:Vk_inv}) always includes flows of higher order invariants. The structure of hierarchy built this way reminds us of that of the Dyson-Schwinger equations for the $n$-point functions; however, the concept here is somewhat different.

Obtaining the flow equations for higher order coefficients is more complicated. For practical purposes, one must close the tower of flow equations anyway, and solve the problem in a restricted coefficient space. One might start with substituting the classical value of $C_k^{(0,1,0,0...)}$ into (\ref{Eq:flow_Uk}); however, we wish to go one step further and derive a flow equation for it, and only use classical values for higher order coefficients. Since classically those are identically zero, the approximation scheme described here actually corresponds to the following approximate chiral invariant expansion of $V_k$:
\bea
\label{Eq:Vk_appr}
V_k(I_1,I_2)=U_k(I_1)+C_k(I_1)\cdot I_2,
\eea
where the notation $C_k^{(0,1,0,0...)}(I_1)\equiv C_k(I_1)$ is used. As described earlier in this subsection, in order to obtain the flow equation for $C_k(I_1)$, one must identify the corresponding coefficient (i.e., of $I_2$) on the right-hand side of the chiral invariant expanded version of (\ref{Eq:flow_Vk}). Since in the background field defined by (\ref{Eq:v0_backg}) $I_2$ vanishes, one needs to extend the background field with more condensate components. Instead of considering the most general scenario (\ref{Eq:gen_backg}), as described earlier in this subsection, due to the approximation (\ref{Eq:Vk_appr}), on top of $v_0$, it is sufficient to introduce only a single, infinitesimal condensate piece. We choose it to be proportional to $T^8$, which refers to the longest diagonal generator (the index notation is just a convention regardless of the actual flavor number $n$, see also the Appendix):
\bea
\bar{M} = v_0 T^0+v_8 T^8.
\label{Eq:v0v8_backg}
\eea
Due to (\ref{Eq:v0v8_backg}), the mass spectrum changes. One has to calculate the modified mass matrices $\partial^2 V_k(I_1,I_2)/\partial s_i \partial s_j|_{v_0,v_8}$ and $\partial^2 V_k(I_1,I_2)/\partial \pi_i \partial \pi_j|_{v_0,v_8}$ and diagonalize them in the $0-8$ subspace. Detailed formulas can be found in the Appendix. After inserting the recalculated masses into the right-hand side of (\ref{Eq:flow_Vk}), one expands the obtained expression around $v_8=0$, while keeping $v_0$ nonzero. Since
\begin{subequations}
\bea
I_1|_{v_0,v_8}&=&\frac{v_0^2+v_8^2}{2}+{\cal O}(v_8^4), \\
I_2|_{v_0,v_8}&=&\frac{v_0^2v_8^2}{n} + {\cal O}(v_8^4),
\eea
\end{subequations}
it is easy to identify these invariants and write the flow equations of $U_k(I_1)$ and $C_k(I_1)$ in an invariant form [higher order terms are negligible based on (\ref{Eq:Vk_appr})]. We note that, in order to get the flows of higher order coefficients, one gradually has to introduce more and more components of the condensate, and eventually use the most general procedure described in the beginning of the subsection. This makes a direct calculation a lot more complicated, but in general there is no restriction to improve the approximation scheme and take into account all invariants. The $I_2$ independent part does not change compared to (\ref{Eq:flow_Uk}), whereas we obtain a new equation from the linear coefficient:
\bea
\label{Eq:flow_Ck}
\partial_k&&C_k(I_1)=k^{d+1}K_d T \sum_{\omega_m}\Bigg[\frac{4(3C_k+2I_1C_k')^2/n}{(\omega_m^2+E_{a_0}^2)^2(\omega_m^2+E_\sigma^2)}\nonumber\\
&&+\frac{128C_k^5I_1^3/n}{(\omega_m^2+E_\pi^2)^3(\omega_m^2+E_{a_0}^2)^3}\nonumber\\
&&+\frac{4C_k\left(4C_k(n^2-3)+(1-4n^2)I_1C_k'\right)/n}{(\omega_m^2+E_{a_0}^2)^3}\nonumber\\
&&+\frac{4\left(3C_kC_k'I_1+4I_1^2C_k'+C_k(3C_k-2C_k''I_1^2)\right)/n}{(\omega_m^2+E_{a_0}^2)(\omega_m^2+E_\sigma^2)^2}\nonumber\\
&&+\frac{64C_k^3I_1^2(C_k-I_1C_k')/n}{(\omega_m^2+E_\pi^2)^2(\omega_m^2+E_{a_0}^2)^3}-\frac{48C_k^2I_1^2C_k'}{(\omega_m^2+E_\pi^2)(\omega_m^2+E_{a_0}^2)^3} \nonumber\\
&&+\frac{6C_k+(1-2n^2)I_1C_k'}{(\omega_m^2+E_{a_0}^2)^2}\frac{1}{I_1}-\frac{6C_k+9I_1C_k'+2I_1^2C_k''}{(\omega_m^2+E_\sigma^2)^2}\frac{1}{I_1}\nonumber\\
&&+\frac{4C_k(6C_k+9I_1C_k'+2I_1^2C_k'')/n}{(\omega_m^2+E_{a_0}^2)(\omega_m^2+E_\sigma^2)^2}\nonumber\\
&&-\frac{2C_k(12C_k+2(1-2n^2)I_1C_k')/n}{(\omega_m^2+E_{a_0}^2)^3}\Bigg].
\eea
All Matsubara sums can be performed analytically with the corresponding formulas presented in the Appendix. Equation (\ref{Eq:flow_Ck}), together with (\ref{Eq:flow_Uk}), can be now solved numerically.

\subsection{Large-$n$ analysis}

The flow equations (\ref{Eq:flow_Uk}) and (\ref{Eq:flow_Ck}) become more simple if $n$ approaches infinity. In order to obtain the appropriate limit, one has to introduce scaling functions instead of $U_k$ and $C_k$. Since the potential $V_k$ itself has to scale with the number of degrees of freedom ($V_k \sim n^2$), in the large-$n$ limit one has
\bea
\label{Eq:largenfunc}
U_k(I_1)=n^2 u_k(i_1), \qquad C_k(I_1)=c_k(i_1)/n,
\eea
where $i_1=I_1/n^2$, since through the condensate, $I_1$ is also scaling with $n^2$. For the derivatives we get
\begin{subequations}
\label{Eq:largenderiv}
\bea
U_k'(I_1)&=&u_k'(i_1), \qquad C_k'(I_1)=c_k'(i_1)/n^3, \\
U_k''(I_1)&=&u_k''(i_1)/n^2, \qquad C_k''(I_1)=c_k''(i_1)/n^5,
\eea
\end{subequations}
showing that they gradually become negligible as the flavor number increases. Inserting (\ref{Eq:largenfunc}) and (\ref{Eq:largenderiv}) into (\ref{Eq:flow_Uk}) and (\ref{Eq:flow_Ck}), at the leading order in the large-$n$ expansion we obtain
\begin{subequations}
\label{Eq:largenflow}
\bea
\partial_k u_k(i_1)&=&k^{d+1}K_d T\sum_{\omega_m} \left[\frac{1}{\omega_m^2+E_{a_0}^2}+\frac{1}{\omega_m^2+E_{\pi}^2}\right], \\
\partial_k c_k(i_1)&=&k^{d+1}K_d T\sum_{\omega_m} \Bigg[\frac{16c_k^2}{\left(\omega_m^2+E_{a_0}^2\right)^3}-\frac{2c_k'}{\left(\omega_m^2+E_{a_0}^2\right)^2}\nonumber\\
&&-\frac{8c_kc_k'i_1}{(\omega_m^2+E_{a_0})^3}+\frac{128c_k^5i_1^3}{(\omega_m^2+E_{\pi}^2)^3(\omega_m^2+E_{a_0}^2)^3} \nonumber\\
&&+\frac{64c_k^3i_1^2(c_k-c_k'i_1)}{(\omega_m^2+E_{\pi}^2)^2(\omega_m^2+E_{a_0}^2)^3}\nonumber\\
&&-\frac{48c_k^2c_k'i_1^2}{(\omega_m^2+E_{\pi}^2)(\omega_m^2+E_{a_0}^2)^3} \Bigg].
\eea
\end{subequations}
Note that now $E_{\pi}^2=k^2+u_k'(i_1)$ and $E_{a_0}^2=E_\pi^2+4i_1c_k(i_1)$. The nature of these equations is similiar to that of (\ref{Eq:flow_Uk}) and (\ref{Eq:flow_Ck}). In the next section we explore how quickly the large-$n$ solution is reached, as $n$ is varied towards infinity.

\subsection{$\beta$ functions}

If one seeks for critical behavior in a system, it is sufficient to search fixed points of the couplings in the dimensionally reduced theory, without the explicit presence of the temperature. Formally it can be obtained by taking the $T\rightarrow \infty$ limit, since in this case the compactified timelike direction disappears (note that after dimensional reduction the left-hand sides of the flow equations are also multiplied by the temperature). In the sums only the $m=0$ term contributes; therefore (\ref{Eq:flow_Uk}) and (\ref{Eq:flow_Ck}) lead to
\begin{subequations}
\label{Eq:flow3d}
\bea
\partial_k&&U_k(I_1)=k^{d+1}K_d \left(\frac{n^2}{k^2+\mu_\pi^2}+\frac{n^2-1}{k^2+\mu_{a_0}^2}+\frac{1}{k^2+\mu_\sigma^2}\right), \nonumber\\
\\
\partial_k&&C_k(I_1)=k^{d+1}K_d \Bigg[\frac{4(3C_k+2I_1C_k')^2/n}{(k^2+\mu_{a_0}^2)^2(k^2+\mu_\sigma^2)}\nonumber\\
&&+\frac{128C_k^5I_1^3/n}{(k^2+\mu_\pi^2)^3(k^2+\mu_{a_0}^2)^3}\nonumber\\
&&+\frac{4C_k\left(4C_k(n^2-3)+(1-4n^2)I_1C_k'\right)/n}{(k^2+\mu_{a_0}^2)^3}\nonumber\\
&&+\frac{4\left(3C_kC_k'I_1+4I_1^2C_k'+C_k(3C_k-2C_k''I_1^2)\right)/n}{(k^2+\mu_{a_0}^2)(k^2+\mu_\sigma^2)^2}\nonumber\\
&&+\frac{64C_k^3I_1^2(C_k-I_1C_k')/n}{(k^2+\mu_\pi^2)^2(k^2+\mu_{a_0}^2)^3}-\frac{48C_k^2I_1^2C_k'}{(k^2+\mu_\pi^2)(k^2+\mu_{a_0}^2)^3} \nonumber\\
&&+\frac{6C_k+(1-2n^2)I_1C_k'}{(k^2+\mu_{a_0}^2)^2}\frac{1}{I_1}-\frac{6C_k+9I_1C_k'+2I_1^2C_k''}{(k^2+\mu_\sigma^2)^2}\frac{1}{I_1}\nonumber\\
&&+\frac{4C_k(6C_k+9I_1C_k'+2I_1^2C_k'')/n}{(k^2+\mu_{a_0}^2)(k^2+\mu_\sigma^2)^2}\nonumber\\
&&-\frac{2C_k(12C_k+2(1-2n^2)I_1C_k')/n}{(k^2+\mu_{a_0}^2)^3}\Bigg].
\eea
\end{subequations}
We note that these equations are generalizations of (39) and (43) of \cite{fukushima10b} for arbitrary $n$. If we further assume that $V_k$ has the form of (\ref{Eq:lag}) but with $k$-dependent couplings,
\bea
V_k(I_1,I_2)=\mu_k^2I_1+\frac{4\pi^2}{3}\left(g_{1,k}+\frac{g_{2,k}}{n}\right)I_1^2+\frac{4\pi^2}{3}g_{2,k}I_2,\nonumber\\
\eea
[note the rescalings $g_{1,k} \longrightarrow 4\pi^2n^2g_{1,k}/3$, $g_{2,k} \longrightarrow 4\pi^2n g_{2,k}/3$ compared to (\ref{Eq:lag})] or equivalently
\begin{subequations}
\bea
U_k(I_1)&=&\mu_k^2 I_1 + \frac{4\pi^2}{3}\left(g_{1,k}+\frac{g_{2,k}}{n}\right)I_1^2, \\
C_k(I_1)&=&\frac{4\pi^2}{3}g_{2,k},
\eea
\end{subequations}
and then the flow equations (\ref{Eq:flow3d}) determine the $\beta$ functions of $g_{1,k}$ and $g_{2,k}$. Introducing scaling variables $\bar{g}_{1,k}=g_{1,k}k^{d-4}$, $\bar{g}_{2,k}=g_{2,k}k^{d-4}$, and setting $\mu^2_k\equiv 0$, in $d=4-\epsilon$ dimensions we get
\begin{subequations}
\label{Eq:beta}
\bea
\beta_1&=&-\epsilon \bar{g}_{1,k}+\frac{n^2+4}{3}\bar{g}_{1,k}^2+\frac{4n}{3}\bar{g}_{1,k}\bar{g}_{2,k}+\bar{g}_{2,k}^2, \\
\beta_2&=&-\epsilon \bar{g}_{2,k}+\frac{2n}{3}\bar{g}_{2,k}^2+2\bar{g}_{1,k}\bar{g}_{2,k},
\eea
\end{subequations}
where the $\beta$ functions are defined as $\beta_1=k\partial \bar{g}_{1,k}/\partial k$, $\beta_2=k\partial \bar{g}_{2,k}/\partial k$ and we are using the approximation $K_d=K_{4-\epsilon}\approx K_4 = (32\pi^2)^{-1}$. Equations (\ref{Eq:beta}) recover the well-known results of $\epsilon$ expansion \cite{pisarski84}, which show that the system has no stable IR fixed point; therefore no second order transition can occur. This argument serves as an indirect evidence of a transition (if it exists) being fluctuation induced, and of first order. In the next section, we shall investigate the solution of the coupled equation system (\ref{Eq:flow_Uk}) and (\ref{Eq:flow_Ck}) to see directly if this argument remains valid for arbitrary $n$. For the sake of completeness, we also list here the flow equation of the mass parameter:
\bea
\label{Eq:massrun}
\frac{\partial \mu_k^2}{\partial k}=-k^{d+1}\frac{(n^2+1)g_{1,k}+2ng_{2,k}}{6(k^2+\mu_k^2)^2}.
\eea
Note that Eqs. (\ref{Eq:beta}), (\ref{Eq:massrun}) are one-loop expressions. To obtain two-loop (or more) results of the perturbative expansion of the $\beta$ functions, one has to go beyond LPA and take into account the momentum dependence of the proper vertices.

\section{Numerical results}

In this section, we present the properties of the numerical solution of the coupled flow equations (\ref{Eq:flow_Uk}) and (\ref{Eq:flow_Ck}). For a given scale $k$, the functions $U_k(I_1)$ and $C_k(I_1)$ are stored on a grid, typically in the region $I_1/\Lambda^2 \in [0,2]$, using a step size of $10^{-3}$. For solving the flow of the equation system (i.e., obtaining the $k$ dependence), we used the Runge-Kutta method, for which a typical step size was chosen to be $\Delta k/\Lambda=10^{-5}$. The field derivatives necessary for the calculation of the masses were obtained with the $7$-point formula, except at the boundaries of the grid, where the $5$- and $3$-point formulas were used.

\begin{figure*}
\begin{center}
\raisebox{0.05cm}{
\includegraphics[keepaspectratio,width=0.345\textwidth,angle=270]{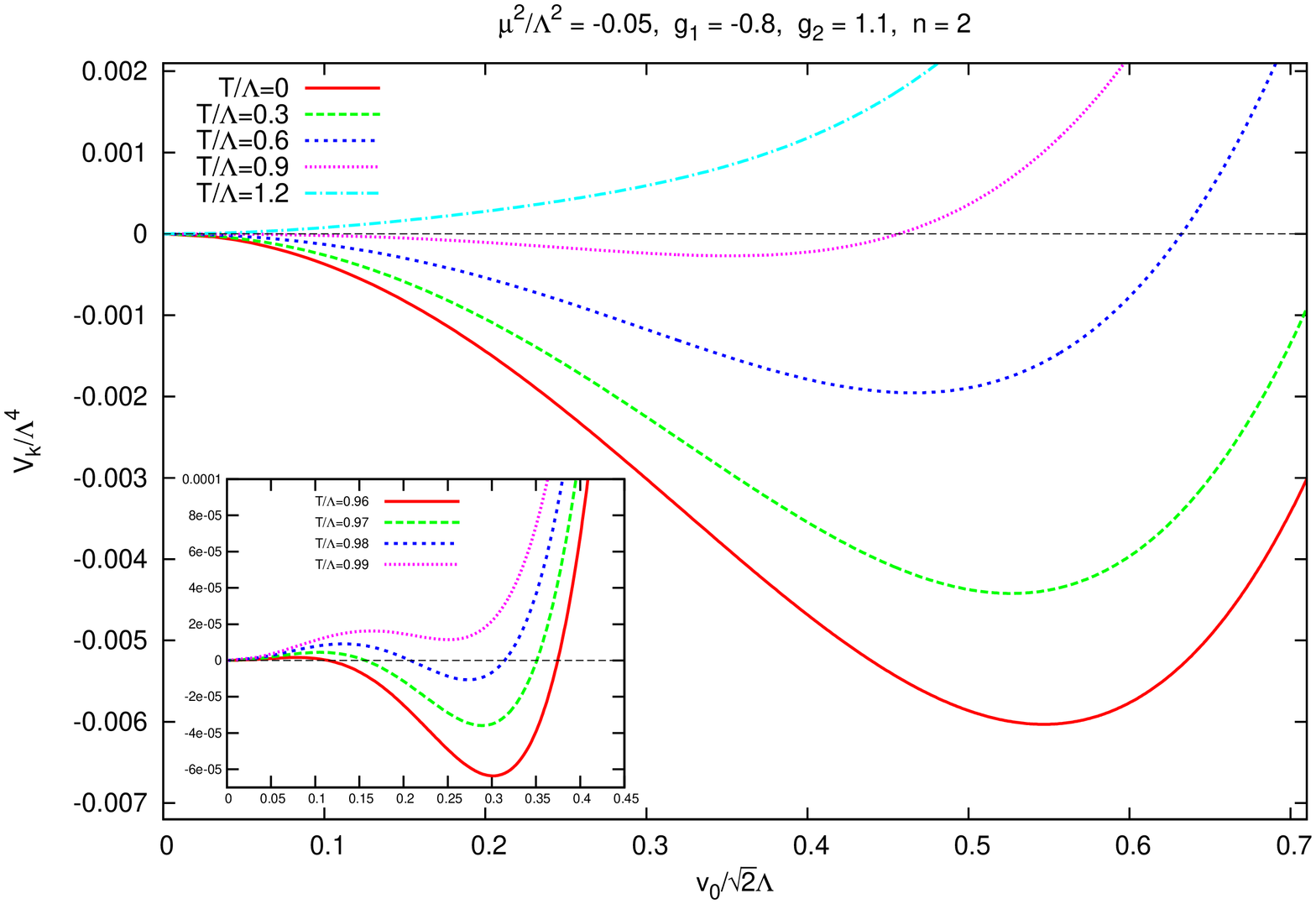}}
\includegraphics[keepaspectratio,width=0.345\textwidth,angle=270]{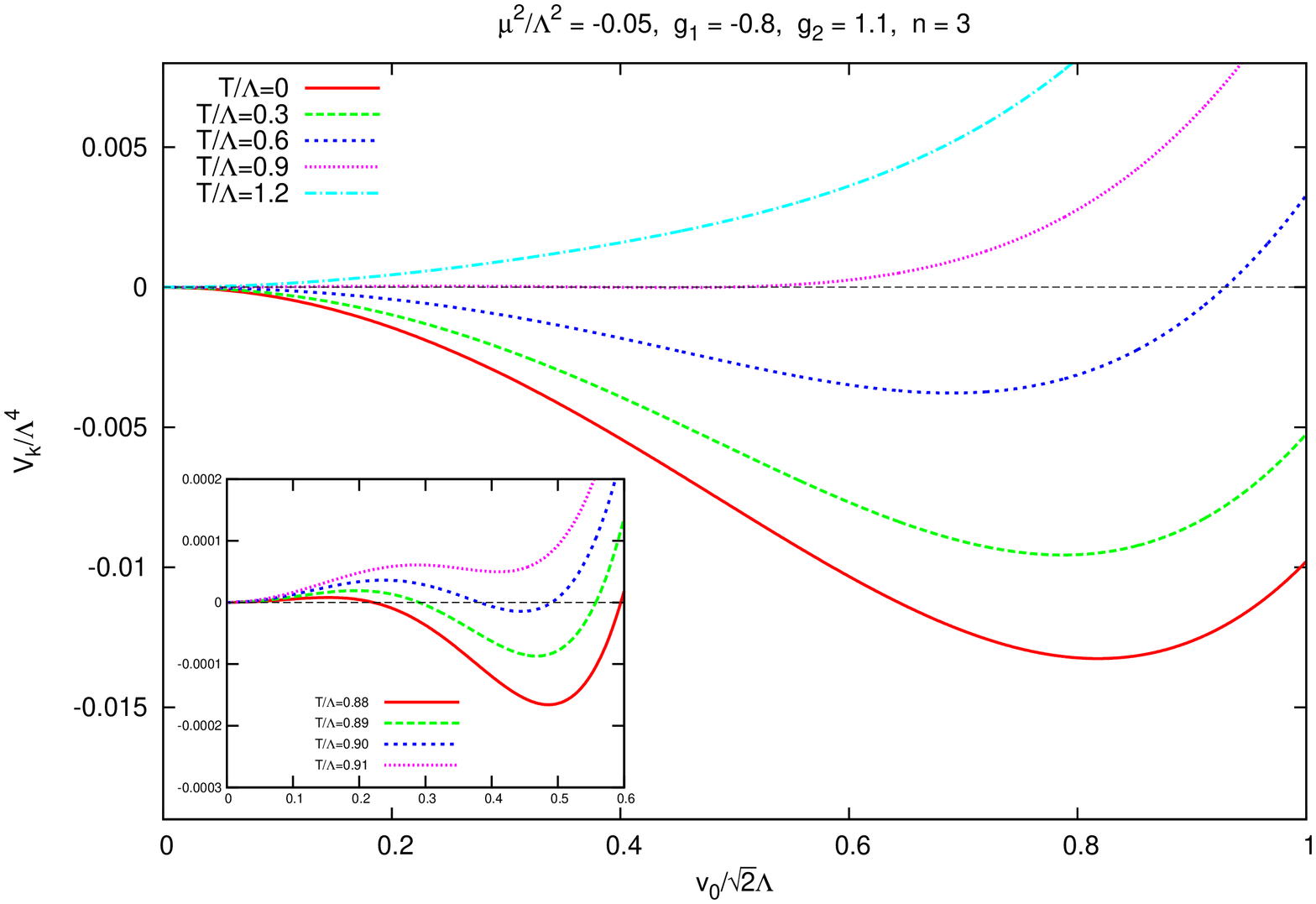}
\caption{Temperature dependence of the effective potential. At each temperature the flow was stopped at $k/\Lambda=0.2$, in order to demonstrate that at intermediate steps the effective potential is not convex, and that the transition temperature can be directly defined. The left figure corresponds to flavor number $n=2$, whereas the right has $n=3$ with the same model parameters.}
\label{fig1}
\end{center}
\end{figure*}

\begin{figure*}
\begin{center}
\raisebox{0.05cm}{
\includegraphics[keepaspectratio,width=0.345\textwidth,angle=270]{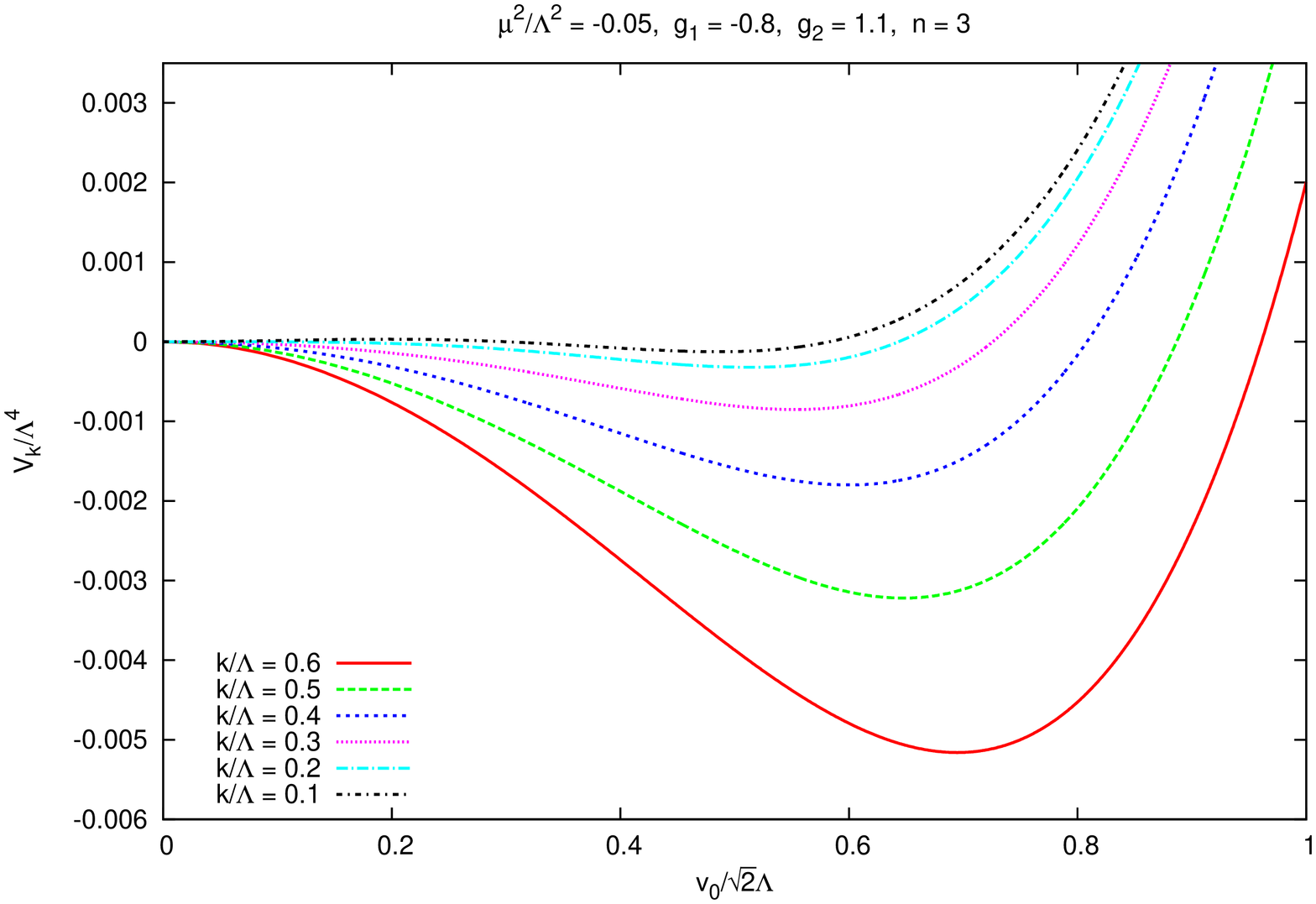}}
\includegraphics[keepaspectratio,width=0.345\textwidth,angle=270]{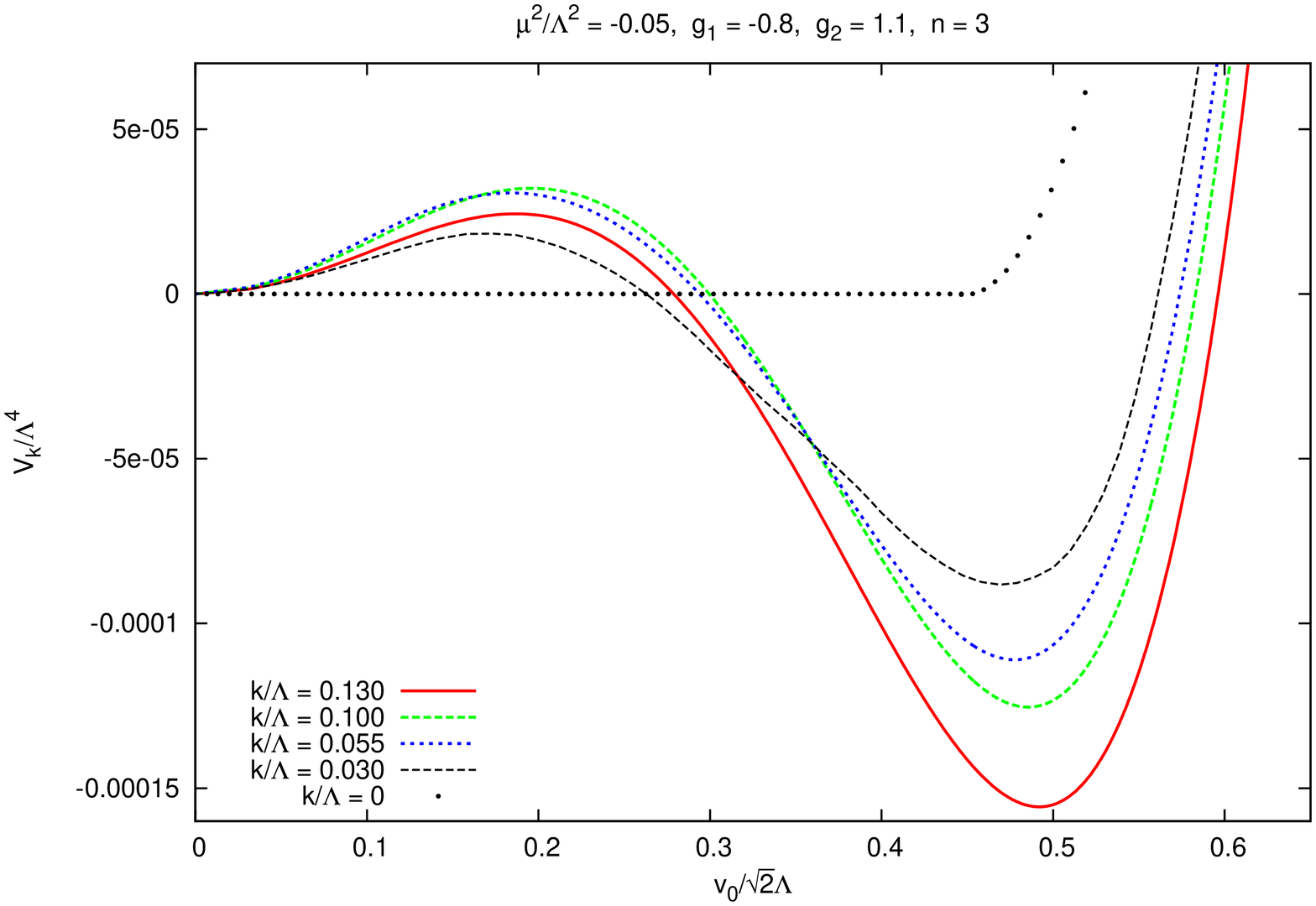}
\caption{Evolution of the effective potential at the critical temperature with respect to scale $k$. The figure on the right shows that at the end of the flow, the potential commences to be convex (the local maximum starts to shrink). The $k/\Lambda=0$, $k/\Lambda=0.03$ curves have not been obtained numerically; they are extrapolated from the larger scale results.}
\label{fig2}
\end{center}
\end{figure*}

\begin{figure*}
\begin{center}
\includegraphics[keepaspectratio,width=0.385\textwidth,angle=270]{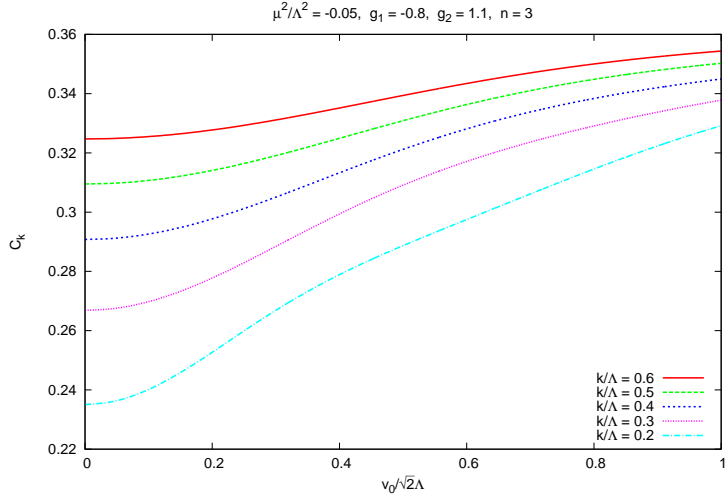}
\caption{Flow of the coefficient function $C_k(I_1)$ in the chiral invariant expansion (\ref{Eq:Vk_appr}). The figure shows that a field independent coupling approximation is crude (i.e., the assumption of $C_k\approx g_{2,k}/n$), since the $C_k(I_1)$ function develops a structure as $k\rightarrow 0$.}
\label{fig3}
\end{center}
\end{figure*}

One of the most important features of the evolution of the effective potential is that it is gradually becoming convex as $k\rightarrow 0$, already confirmed by several studies in the literature \cite{berges02,fukushima10b,litim94,litim96}. This phenomenon is not surprising, since unlike perturbation theory, FRG formalism does obey the convexity of the effective potential, which can be directly shown from the flow equation (\ref{Eq:flow}), and which is respected by LPA. Since the obtained full quantum effective action is convex, the question arises that, if the transition is of first order, how one can determine the transition temperature and the discontinuity of the order parameter. The answer lies in the fact that even though at $k=0$ the potential is convex, it is not true for any nonzero scales \cite{ringwald90} (see typical examples in Fig. \ref{fig1}). Depending on the physical system in question, it might be appropriate to stop the RG flow before reaching $k=0$, since the inverse of the flow parameter may be identified with a finite correlation length, serving as a physical infrared cutoff and leading $V_k$ to the corresponding coarse grained free energy density. In this case, the transition temperature $T_c$ can be directly defined by requiring both minima to have the same energy. Mathematically, $T_c$ and the discontinuity of the order parameter are defined through the equations
\bea
\label{Eq:Tcdet}
V_k(0;T_c)-V_k(v_0^*;T_c)=0, \quad \frac{\partial V_k}{\partial v_0}(v_0^*;T_c)=0.
\eea
The second equation implies that $v_0^*$ is the discontinuity of the order parameter at the transition point (note that $\sqrt{2I_1|_{v_0}}=v_0$). These quantities depend on the actual scale $k$; nevertheless we shall obtain a prediction independently of the infrared cutoff of the system. Considering this purpose, the critical temperature and the discontinuity of the order parameter are defined as the following limits: $\lim_{k\rightarrow 0} T_c(k)$ and $\lim_{k\rightarrow 0} v_0^*(k)$, respectively. Since from a numerical point of view, it requires an unreasonably long computational time to reach $k=0$ itself, we extrapolate our results from nonzero scales to $k=0$. For both curves $T_c(k)$ and $v_0^*(k)$, a function of the form $f(k)=a+b\cdot k^c$ has been identified as a good approximation for fitting, leading to reliable values of $T_c(k=0)$ and $v_0^*(k=0)$. A typical evolution of the effective potential can be seen in Fig. \ref{fig2}, together with the flow of the $C_k(I_1)$ coefficient function of the invariant expansion (\ref{Eq:Vk_appr}) in Fig. \ref{fig3}. The latter shows that approximating $C_k(I_1)$ with a constant [i.e., $C_k(I_1)\approx g_{2,k}/n$] is rather crude, given that the function starts to develop a structure as $k\rightarrow 0$. 

\begin{figure*}
\begin{center}
\raisebox{0.05cm}{
\includegraphics[keepaspectratio,width=0.335\textwidth,angle=270]{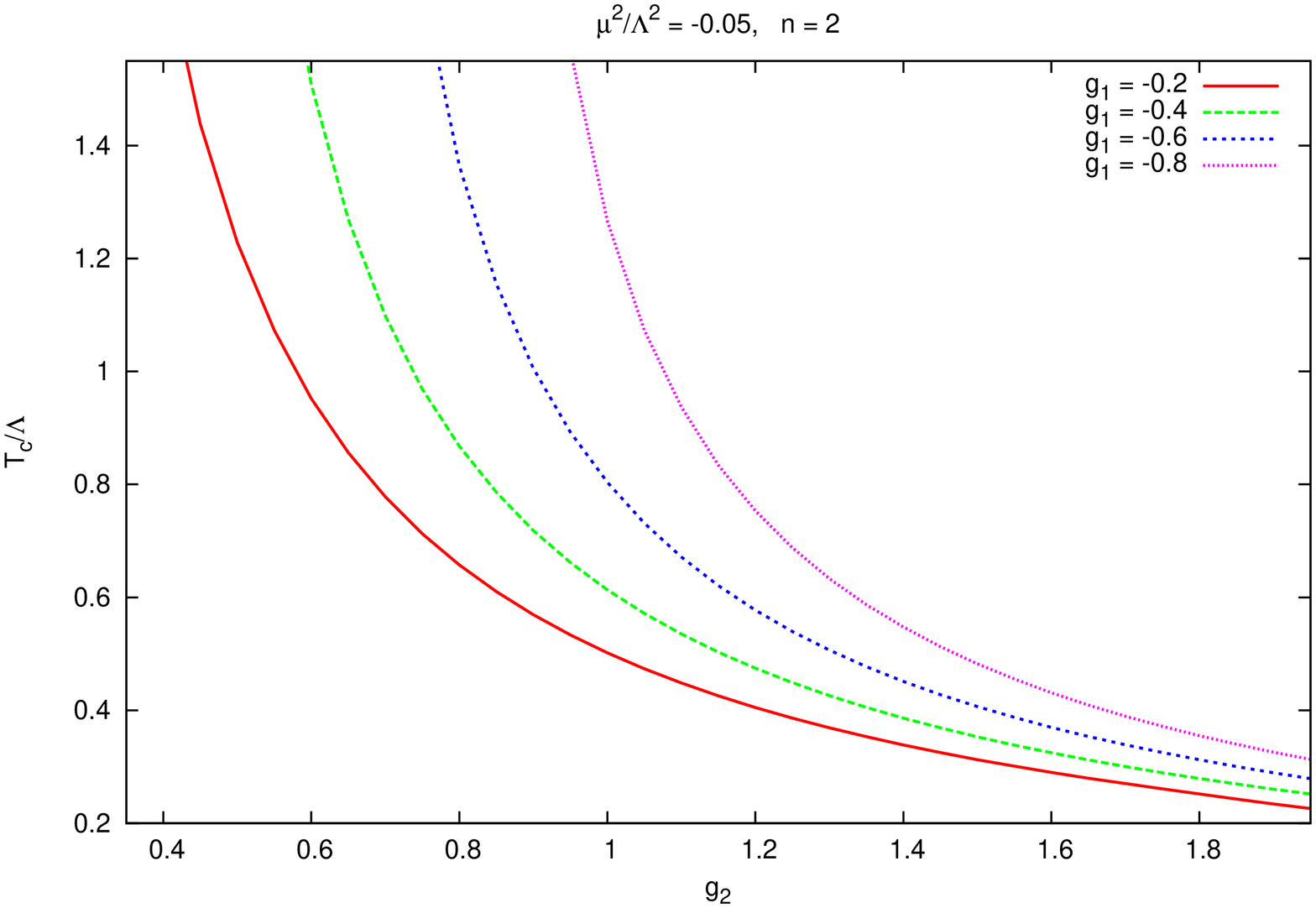}}
\includegraphics[keepaspectratio,width=0.335\textwidth,angle=270]{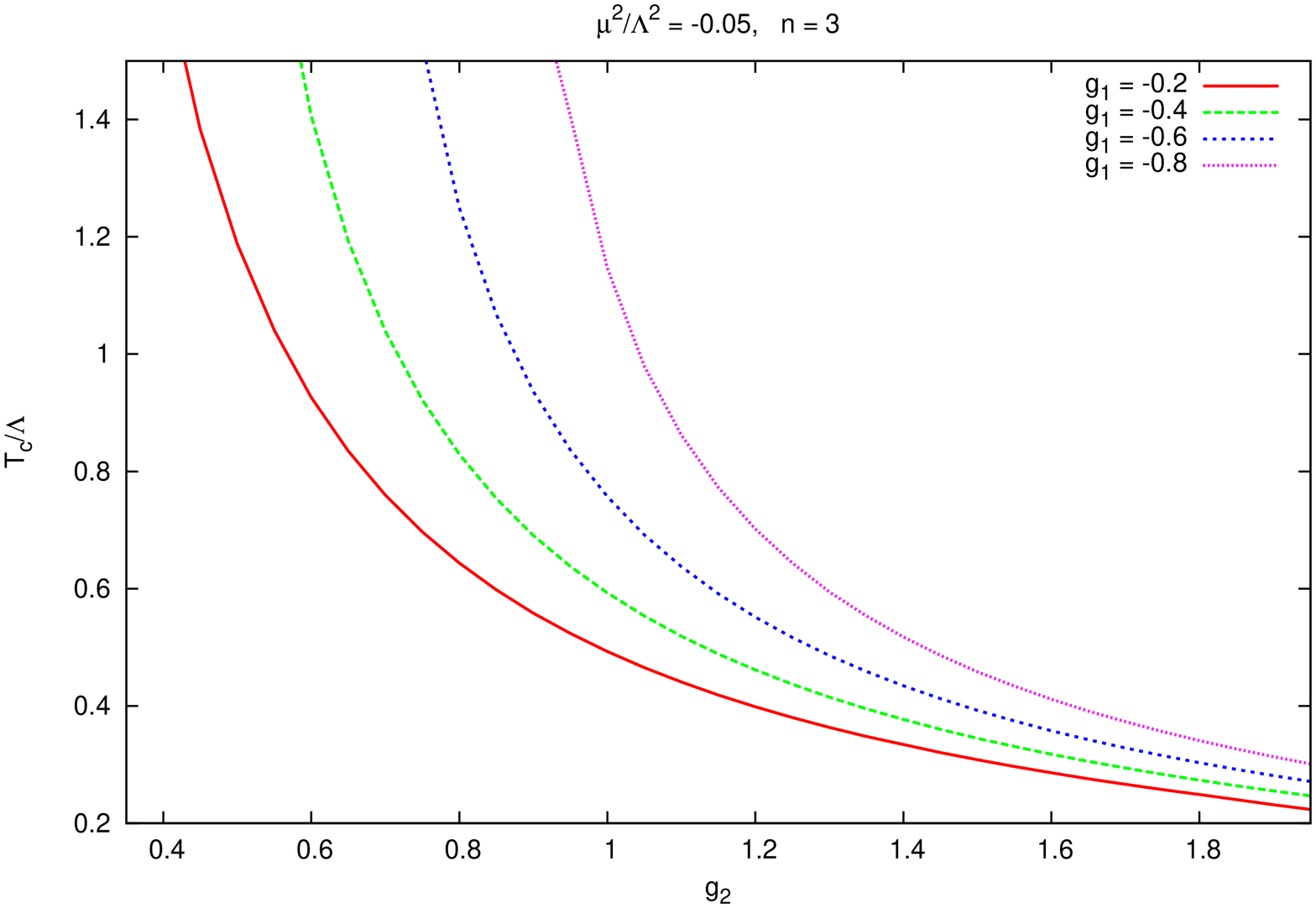}
\includegraphics[keepaspectratio,width=0.335\textwidth,angle=270]{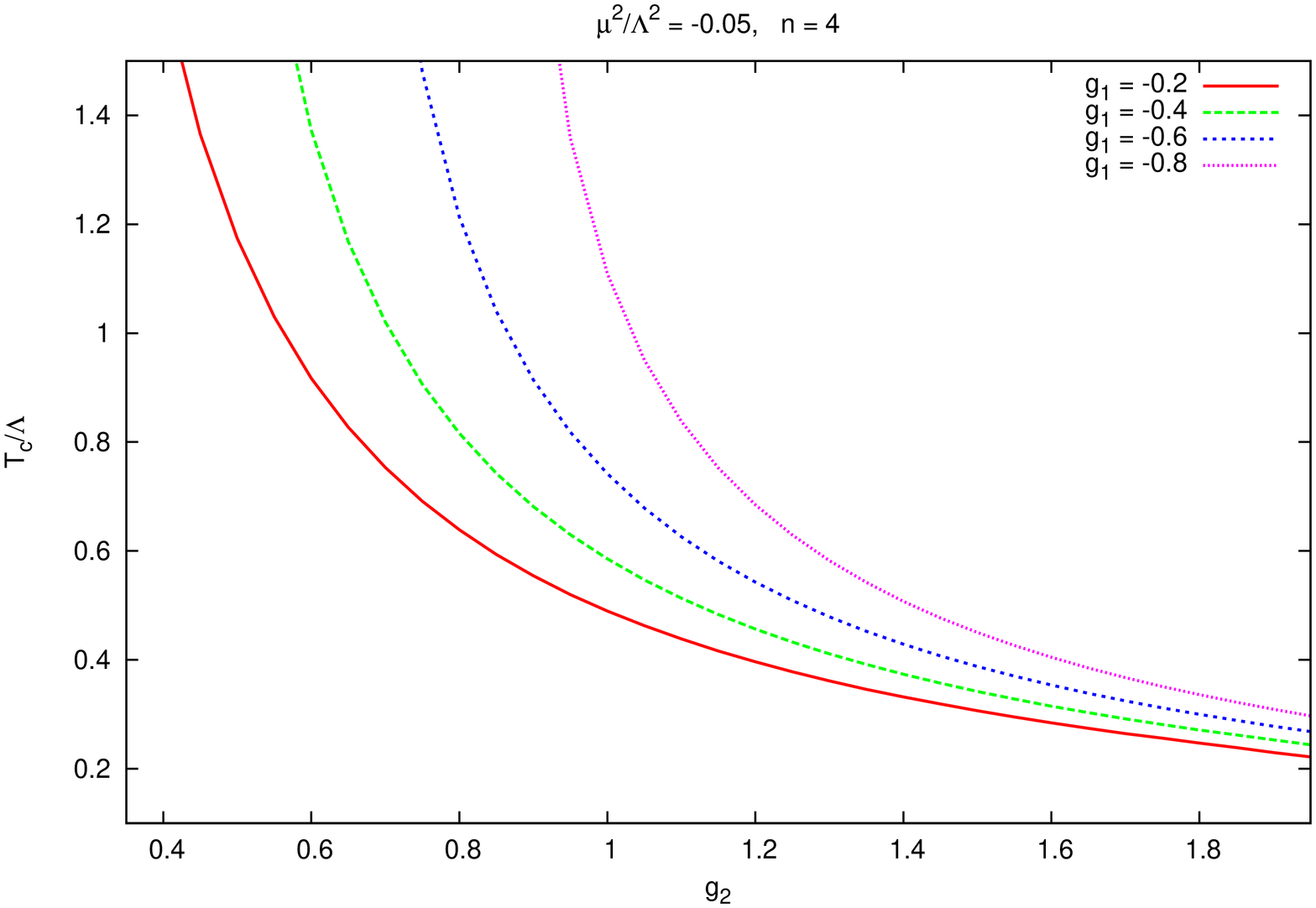}
\includegraphics[keepaspectratio,width=0.335\textwidth,angle=270]{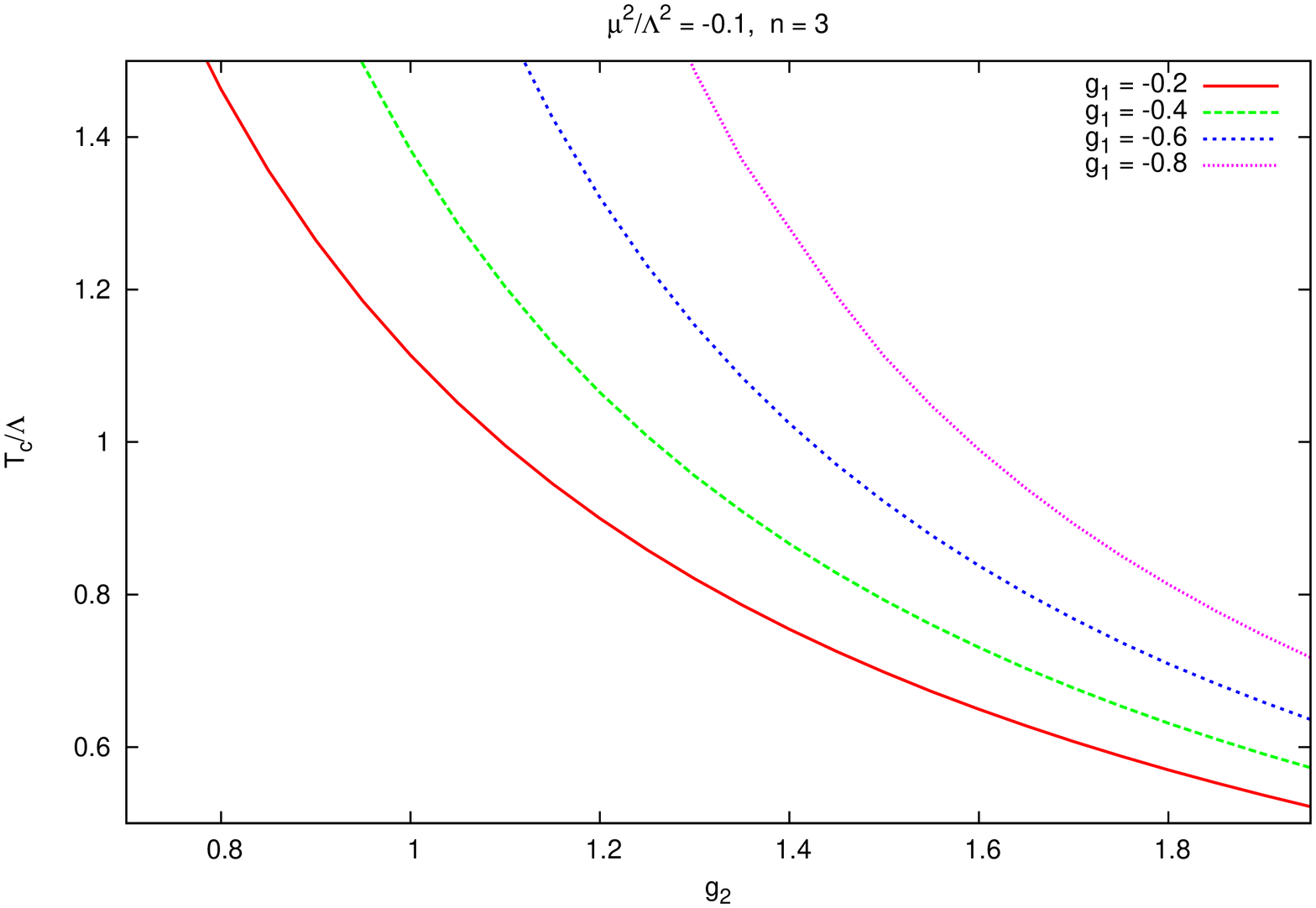}
\caption{Comparison of the critical temperature for several flavor numbers and model parameters.}
\label{fig4}
\end{center}
\end{figure*}

\begin{figure*}
\begin{center}
\raisebox{0.05cm}{
\includegraphics[keepaspectratio,width=0.335\textwidth,angle=270]{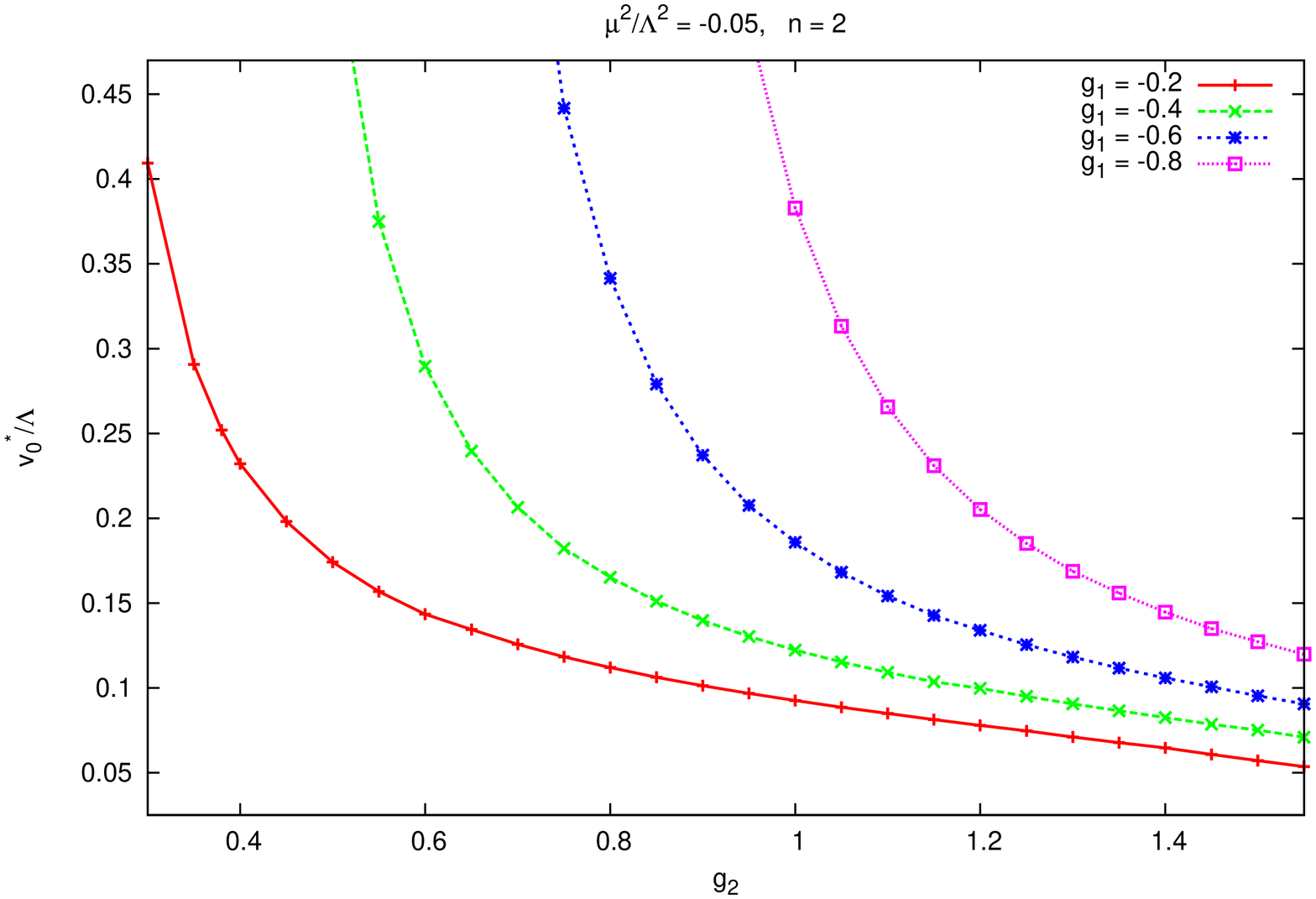}}
\includegraphics[keepaspectratio,width=0.335\textwidth,angle=270]{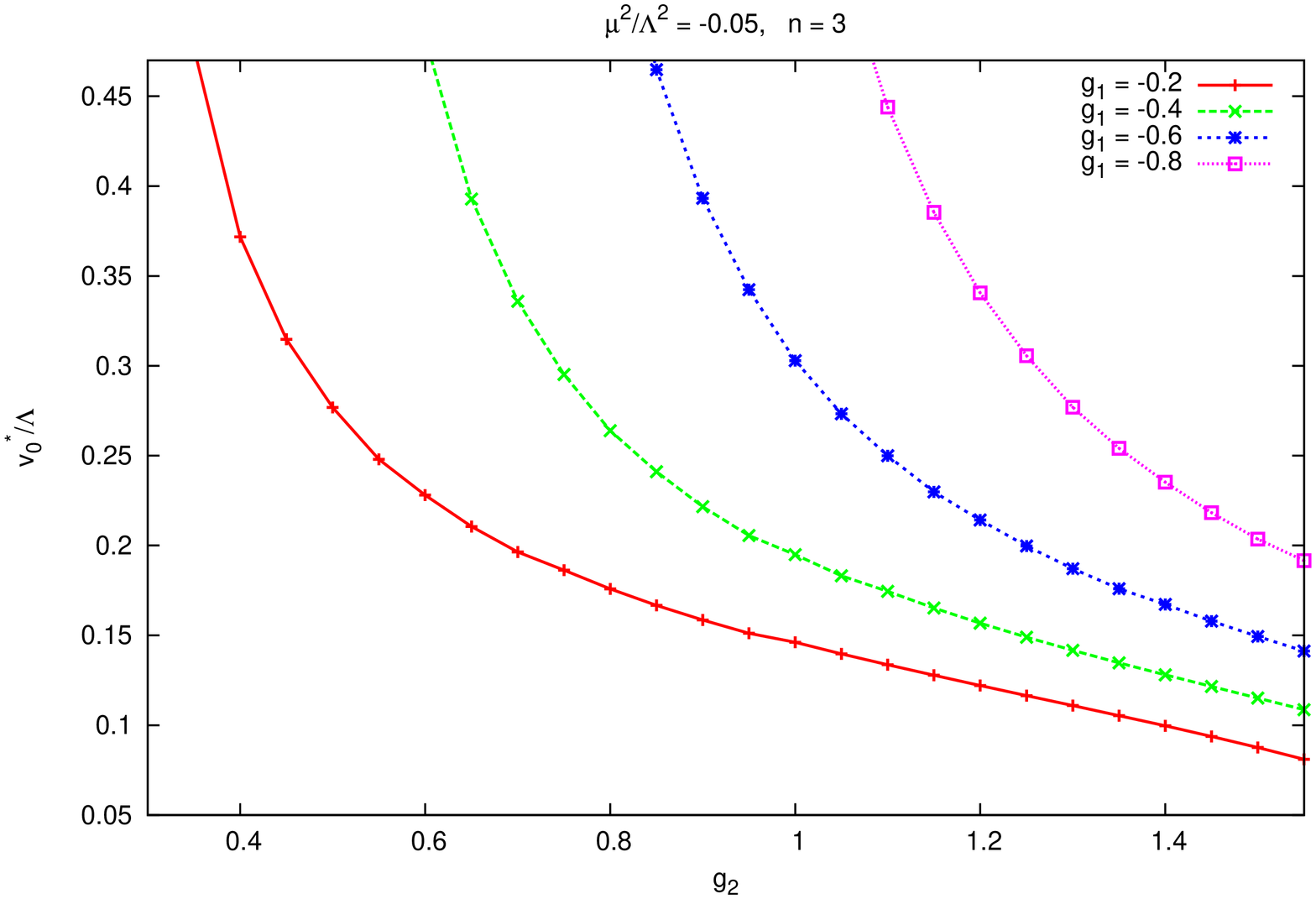}
\includegraphics[keepaspectratio,width=0.335\textwidth,angle=270]{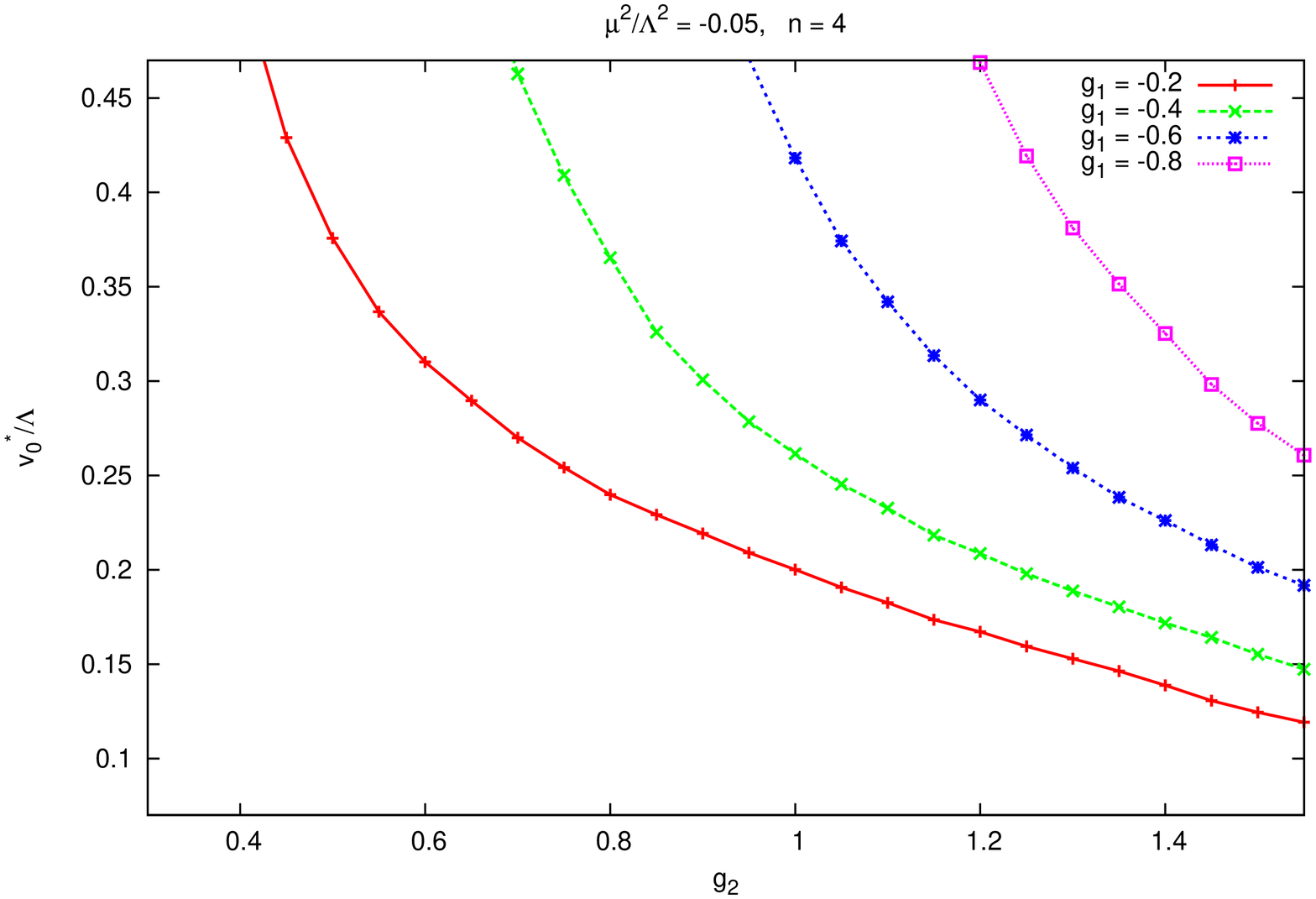}
\includegraphics[keepaspectratio,width=0.335\textwidth,angle=270]{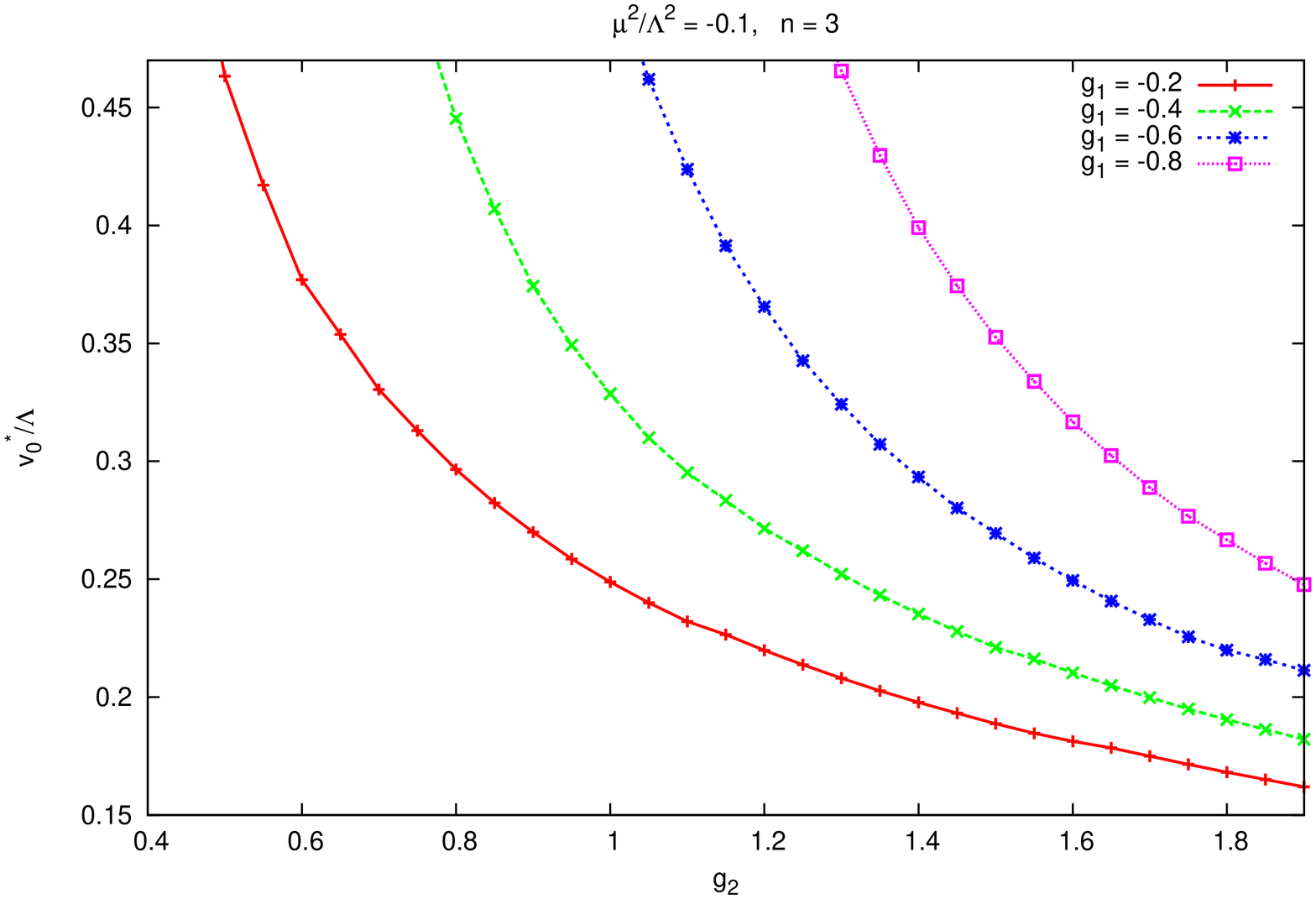}
\caption{Comparison of the discontinuity of the order parameter at the transition point for several flavor numbers and model parameters.}
\label{fig5}
\end{center}
\end{figure*}

Now we review the results concerning the transition temperature $T_c$ and the $v_0^*$ discontinuity of the order parameter. The results for several flavor numbers and model parameters are shown in Figs. {\ref{fig4}} and {\ref{fig5}}, respectively. First we observe that, for a given mass parameter, smaller $g_1$ leads to smaller $T_c$, as a function of $g_2$. Secondly, we see that increasing the flavor number does not change considerably the dependence of the critical temperature with respect to any of the couplings $g_1, g_2$. Note that had we not included the proper $n$-scalings of these couplings from the beginning in (\ref{Eq:lag}), this statement would not be true. On the contrary, if we increase the absolute value of the mass parameter, the $T_c=T_c(g_2)$ curves do change; for fixed $g_1, g_2$, the transition temperature monotonically grows, as expected, since at zero temperature, the symmetry breaking minimum gets deeper. Concerning $v_0^*$, it is more sensitive to the flavor number than the transition temperature. For a given set of coupling constants, the higher $n$ is, the higher the discontinuity of the order parameter will be, and the same tendency is observed regarding the mass parameter. Otherwise, the curves are very similar to that of the transition temperature: at fixed $g_1$, the $v_0^* = v_0^*(g_2)$ function is decreasing. Note that stability requires $g_1+g_2>0$; therefore if $g_1<0$, at $g_2=|g_1|$, both $T_c$ and $v_0^*$ diverge, as they should.

\begin{figure*}
\begin{center}
\raisebox{0.05cm}{
\includegraphics[keepaspectratio,width=0.395\textwidth,angle=270]{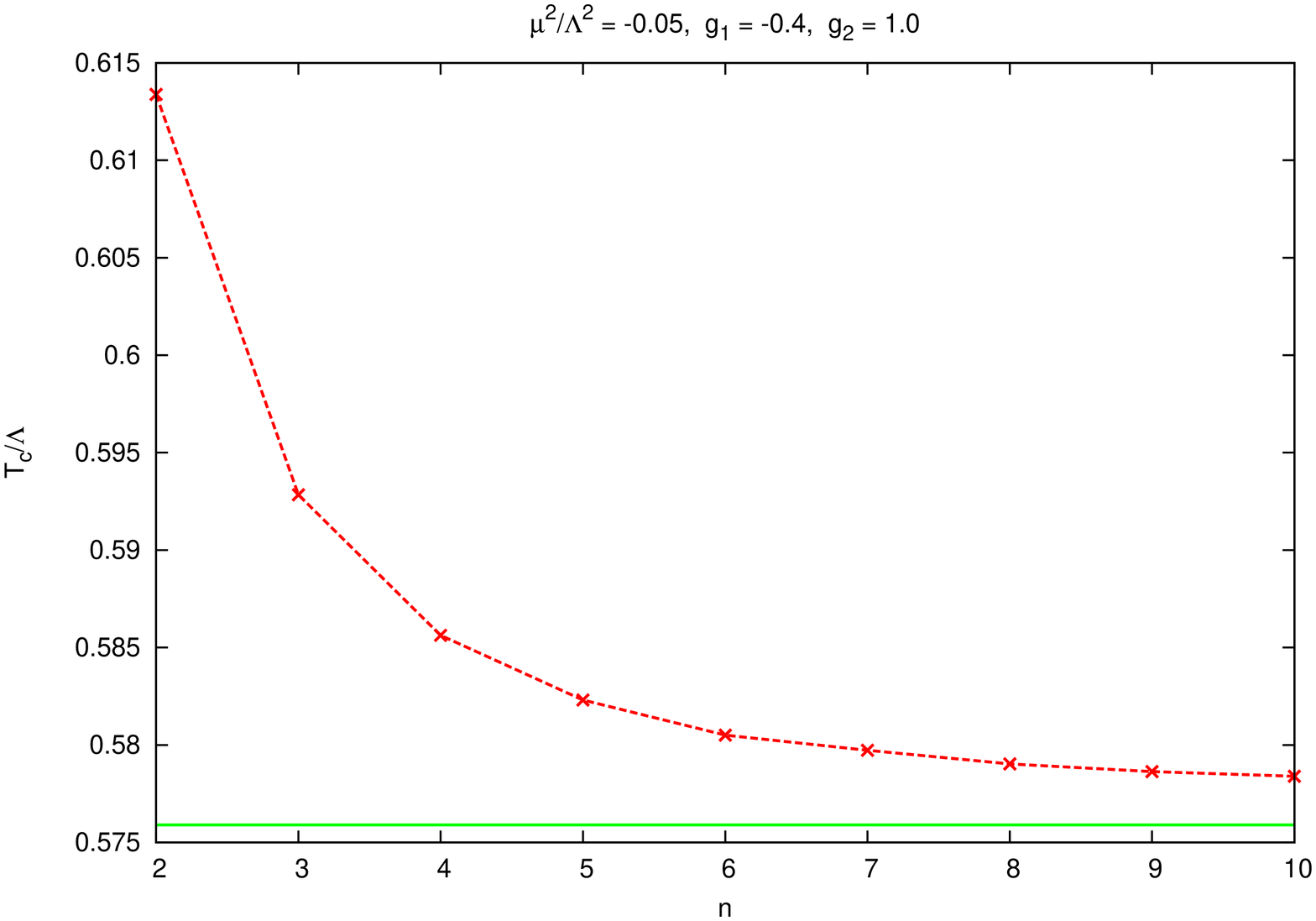}}
\caption{Critical temperature for increasing flavor number in comparison with the large-$n$ result (straight line).}
\label{fig6}
\end{center}
\end{figure*}

Finally let us compare the large-$n$ solution [i.e. solution of Eqs. (\ref{Eq:largenflow})] with the results for finite flavor numbers. In Fig. \ref{fig6}, the critical temperature for increasing flavor numbers is demonstrated, together with the convergence to the large-$n$ value. What we observe is that even at $n=3$, the difference between the large-$n$ and the $n=3$ results is less than 3\% for our particular set of parameters. The same results apply for the discontinuity $v_0^*$ (note that it scales as $v_0^*\sim n$). This suggests that in the $U(n)\times U(n)$ models, even a leading order result in the large-$n$ expansion might be valuable for the phenomenologically most important $n=3$ case. 

\section{Conclusions}

We have used the functional renormalization group method equipped with the local potential approximation to analyze the phase transition in the family of $U(n)\times U(n)$ symmetric scalar theories. Based on the expectation that the system undergoes a first order transition, we have worked in $3+1$ dimensions, with an explicit presence of the temperature, since dimensional reduction can only take place if the system is close to criticality. We have proposed to calculate the effective potential via a chiral invariant expansion, leading to field dependent coefficients of each term in the expansion, for which coupled functional flow equations have been derived. The hierarchy they form is similar but not identical to that of the Dyson-Schwinger equations for the $n$-point functions. With approximating the chiral invariant expansion with field independent couplings, the resulting flow equations reproduced the well-known $\epsilon$-expansion results of the $\beta$ functions at one-loop level. This showed that at least within this approximation, there was no stable infrared fixed point for any flavor number. Note that for $n=2$, however, there are other methods showing a second order transition in the system \cite{pelissetto13,nakayama14}.

The flow equations were solved numerically on a grid, and regardless of the flavor number and directions in the parameter space, first order transitions have been observed. The properties of the transition have been mapped in a large region of the parameter space, extended with a large-$n$ analysis. It has been shown that the effective potential is getting convex in the IR limit; therefore the transition temperature and the discontinuity can be defined only as appropriate limits of the flows. We have also solved the coupled flow equations in the leading order of the large-$n$ expansion and found that even at lower flavor numbers (i.e., $n \gtrsim 3$), the obtained results for the critical temperature and discontinuity are already close to those obtained for definite flavor numbers. This suggests that the large-$n$ expansion is quite robust, and combining it with more sophisticated approximations of the FRG flows could give reliable results even for lower flavor numbers.

The method of the chiral invariant expansion can be systematically improved by taking into account more invariant tensors. This would be particularly interesting for $n=3$, where only three invariant tensors are present, but the role of the cubic one in $M^\dagger M$ has not been investigated before. Furthermore, the method can be directly extended to theories with $U_A(1)$ anomaly and/or finite quark masses included. One may check the temperature dependence of the anomaly around the critical point, investigate the order of the transition with respect to light and heavy quark masses, and build phenomenology upon the scheme presented here. These represent future studies to be reported.

\section*{Acknowledgements}
The author thanks T. Hatsuda, K. Kamikado, A. Patk\'os and Zs. Sz\'ep for useful discussions and suggestions. This work was supported by the Foreign Postdoctoral Research Program of RIKEN.
\newline

\makeatletter
\@addtoreset{equation}{section}
\makeatother 
\newpage
\renewcommand{\theequation}{A\arabic{equation}} 

\appendix 
\section{Properties of the U(n) group}   

The $U(n)$ group has $n$ diagonal and $n(n-1)$ nondiagonal independent generators leading to $n^2$ in total. They are traceless and normalized as $\Tr(T^aT^b)=\delta^{ab}/2$. The diagonal ones read as
\bea
T^{(0)}&=&{\frac{1}{\sqrt{2n}}}
\left( \begin{array}{cccc}
1 & & & \\
& 1 & & \\
& & ... & \\
& & & 1 \\
\end{array} \right), \nonumber\\
T^{(1)}&=&\frac12
\left( \begin{array}{cccc}
1 & & & \\
& -1 & & \\
& & 0 & \\
& & & ... \\
\end{array} \right),\nonumber\\
&...&\nonumber\\
T^{(n-1)}&=&\frac{1}{\sqrt{2n(n-1)}}
\left( \begin{array}{cccc}
1 & & & \\
& 1 & & \\
& & ... & \\
& & & -(n-1) \\
\end{array} \right).
\eea
We note that throughout the paper we follow the notation $T^{(n-1)}\equiv T^8$, irrespective of the actual value of the flavor number $n$.
The nondiagonal generators form two groups, as generalizations of $(x)$- and $(y)$-type of Pauli matrices in the following sense:
\begin{subequations}
\label{app-nondiag}
\bea
\big(T^{(x,jk)}\big)_{ab}&=&\frac12(\delta_{ak}\delta_{bj}+\delta_{aj}\delta_{bk}), \\
\big(T^{(y,jk)}\big)_{ab}&=&\frac{i}{2}(\delta_{ak}\delta_{bj}-\delta_{aj}\delta_{bk}),
\eea
\end{subequations}
where the obvious index notations $(x,jk)$ and $(y,jk)$ ($j<k$) have been introduced. The $d_{abc}$ symmetric, and the $f_{abc}$ antisymmetric structure constants of the $U(n)$ Lie algebra are defined via the relation
\bea
T^aT^b=\frac12(d_{abc}+if_{abc})T^c.
\eea
The following identities are useful for obtaining the structure constants:
\begin{subequations}
\label{app-struct}
\bea
[T^a,T^b]&=&if_{abc}T^c \quad \Longrightarrow \quad \Tr\big[[T^a,T^b]T^c\big]=if_{abc}/2, \nonumber\\
\\
\{T^a,T^b\}&=&d_{abc}T^c \quad \Longrightarrow \quad \Tr \big[\{T^a,T^b\}T^c\big]=d_{abc}/2, \nonumber\\
\eea
\end{subequations}
where $[.,.]$ and $\{.,.\}$ refer to commutation and anticommutation, respectively. Alternatively, we can write 
\bea
f_{abc}=4\Im \Tr(T^aT^bT^c), \qquad d_{abc}=4\Re \Tr(T^aT^bT^c). \nonumber\\
\eea
For determining various invariants and their derivatives, the following structure constants are necessary to be computed:
\begin{widetext}
\bea
\label{Eq:structcon}
d^{0ij}&=&\sqrt{\frac{2}{n}}\delta^{ij}, \qquad d^{888}=(2-n)\sqrt{\frac{2}{n(n-1)}}, \qquad d^{8ij\neq 0,8}=\begin{cases} (2-n)\sqrt{\frac{1}{2n(n-1)}}\delta^{ij} , \hspace{0.1cm} \ifff \hspace{0.1cm} i,j \in \{(x,jn),(y,jn)\} \\ \sqrt{\frac{2}{n(n-1)}}\delta^{ij}, \hspace{1.3cm} \elsee \end{cases},\nonumber\\
d^{i88}&=&\sqrt{\frac{2}{n}}\delta^{i0}+d^{888}\delta^{i8}, \qquad f^{u8v}=\sqrt{\frac{n}{2(n-1)}}\left(\delta_{u,(y,jn)}\delta_{v,(x,jn)}-\delta_{u,(x,jn)}\delta_{v,(y,jn)}\right).
\eea
\makeatletter
\@addtoreset{equation}{section}
\makeatother 

\renewcommand{\theequation}{B\arabic{equation}} 

\section{Mass matrices and group invariants}  
 
Here we list the elements of the mass matrices $\mu_{s,ab}^2$ and $\mu_{\pi,ab}^2$, which are necessary to be computed in order to obtain the flow equation (\ref{Eq:flow_Ck}). Note that the effective potential is approximated as $V_k=U_k(I_1)+C_k(I_1)\cdot I_2$.
\begin{subequations}
\bea
\mu_{s,ab}^2\equiv \frac{\partial^2 V_k}{\partial s^a \partial s^b}\bigg|_{v_0,v_8}&=&\frac{\partial I_1}{\partial s^a}\frac{\partial I_1}{\partial s^b}\bigg|_{v_0,v_8}\Big(U_k''(I_1)+C_k''(I_1)\cdot I_2\Big)+\frac{\partial^2 I_1}{\partial s^a\partial s^b}\bigg|_{v_0,v_8}\Big(U_k'(I_1)+C_k'(I_1)\cdot I_2\Big)\nonumber\\
&+&\Big(\frac{\partial I_1}{\partial s^a}\frac{\partial I_2}{\partial s^b}+\frac{\partial I_1}{\partial s^b}\frac{\partial I_2}{\partial s^a}\Big)\bigg|_{v_0,v_8}C_k'(I_1)+\frac{\partial^2 I_2}{\partial s^a s^b}\bigg|_{v_0,v_8}C_k(I_1), \\
\mu_{\pi,ab}^2\equiv \frac{\partial^2 V_k}{\partial \pi^a \partial \pi^b}\bigg|_{v_0,v_8}&=&\frac{\partial I_1}{\partial \pi^a}\frac{\partial I_1}{\partial \pi^b}\bigg|_{v_0,v_8}\Big(U_k''(I_1)+C_k''(I_1)\cdot I_2\Big)+\frac{\partial^2 I_1}{\partial \pi^a\partial \pi^b}\bigg|_{v_0,v_8}\Big(U_k'(I_1)+C_k'(I_1)\cdot I_2\Big)\nonumber\\
&+&\Big(\frac{\partial I_1}{\partial \pi^a}\frac{\partial I_2}{\partial \pi^b}+\frac{\partial I_1}{\partial \pi^b}\frac{\partial I_2}{\partial \pi^a}\Big)\bigg|_{v_0,v_8}C_k'(I_1)+\frac{\partial^2 I_2}{\partial \pi^a \pi^b}\bigg|_{v_0,v_8}C_k(I_1).
\eea
\end{subequations}
The necessary invariants and their derivatives in the background field $\bar{M}=v_0T^0+v_8T^8$ up to ${\cal O}(v_8^2)$ (note that calculating higher order contributions are negligible in our scheme) are as follows:
\begin{subequations}
\bea
I_1|_{v_0,v_8}&=&\frac{v_0^2+v_8^2}{2}, \hspace{6.1cm} I_2|_{v_0,v_8}=v_0^2v_8^2/n, \\
\frac{\partial I_1}{\partial s^a}\bigg|_{v_0,v_8}&=&v_0\delta^{a0}+v_8\delta^{a8}, \qquad \qquad \qquad \qquad \qquad \qquad \qquad \frac{\partial I_1}{\partial \pi^a}\bigg|_{v_0,v_8}=0, \\
\frac{\partial I_2}{\partial s^a}\bigg|_{v_0,v_8}&=&\frac{2v_0v_8^2}{n}\delta_{a0}+\left(\frac{2v_0^2v_8}{n}-\frac{3(n-2)v_0v_8^2}{n\sqrt{n-1}}\right)\delta_{a8}, \qquad \frac{\partial I_2}{\partial \pi^a}\bigg|_{v_0,v_8}=0, \\
\frac{\partial^2 I_1}{\partial s^a \partial s^b}\bigg|_{v_0,v_8}&=&\delta^{ab}, \qquad \qquad \qquad \qquad \qquad \qquad \qquad \qquad \quad \frac{\partial^2 I_1}{\partial \pi^a \partial \pi^b}\bigg|_{v_0,v_8}=\delta^{ab}, \\
\frac{\partial^2 I_2}{\partial s^a s^b}\bigg|_{v_0,v_8}&=&
\begin{cases}
\frac{2}{n}v_8^2,  \hspace{4.5cm} \ife \hspace{0.1cm} a=b=0\\
-\frac{3(n-2)}{n\sqrt{n-1}}v_8^2+\frac{4}{n}v_0v_8,  \hspace{2.15cm} \ife \hspace{0.1cm} a=0,\hspace{0.1cm} b=8 \hspace{0.1cm} \orr \hspace{0.1cm} a=8,\hspace{0.1cm} b=0\\
\frac{2}{n}v_0^2+\frac{3(n-2)^2}{n(n-1)}v_8^2-\frac{6(n-2)}{n\sqrt{n-1}}v_0v_8,  \hspace{0.55cm} \ife \hspace{0.1cm} a=b=8\\
\frac{2}{n}v_0^2+\frac{4-n}{n(n-1)}v_8^2+\frac{6}{n\sqrt{n-1}}v_0v_8, \hspace{0.67cm} \ife \hspace{0.1cm} a=b\in \pion\\
\frac{2}{n}v_0^2+\frac{(n-2)^2}{n(n-1)}v_8^2-\frac{3(n-2)}{n\sqrt{n-1}}v_0v_8, \hspace{0.68cm} \ife \hspace{0.1cm} a=b\in \kaon\\
0, \hspace{4.99cm}  \els\\
\end{cases}\\
\frac{\partial^2 I_2}{\partial \pi^a \pi^b}\bigg|_{v_0,v_8}&=&
\begin{cases}
0, \hspace{4.95cm} \ife \hspace{0.1cm} a=b=0\\
-\frac{n-2}{n\sqrt{n-1}}v_8^2+\frac{2}{n}v_0v_8, \hspace{2.15cm} \ife \hspace{0.1cm} a=0,\hspace{0.1cm} b=8 \hspace{0.1cm} \orr \hspace{0.1cm} a=8,\hspace{0.1cm} b=0\\
\frac{(n-2)^2}{n(n-1)}v_8^2-2\frac{n-2}{n\sqrt{n-1}}v_0v_8, \hspace{1.51cm} \ife \hspace{0.1cm} a=b=8\\
-\frac{n-2}{n(n-1)}v_8^2+\frac{2}{n\sqrt{n-1}}v_0v_8, \hspace{1.42cm} \ife \hspace{0.1cm} a=b\in \pion\\
\frac{n^2-2n+2}{n(n-1)}v_8^2-\frac{n-2}{n\sqrt{n-1}}v_0v_8, \hspace{1.43cm} \ife \hspace{0.1cm} a=b\in \kaon\\
0, \hspace{4.98cm}  \els.\\
\end{cases}
\eea
\end{subequations}
An index ``a'' is denoted as kaon type, if $a=(x,jn)$ or $a=(y,jn)$, where $j=0,...n-1$. If ``a'' is not kaon type and $a\neq 0,8$, then it is called pion type. 
\renewcommand{\theequation}{C\arabic{equation}} 

\section{Matsubara sums}

The right-hand side of the flow equations (\ref{Eq:flow_Uk}) and (\ref{Eq:flow_Ck}) contains several different types of Matsubara sums. Using the notation
\bea
S(i,j)=\sum_{\omega_m} \frac{1}{(\omega_m^2+E_1^2)^i (\omega_m^2+E_2^2)^j},
\eea
the sums to be calculated are $S(1,0)$, $S(2,0)$, $S(3,0)$, $S(2,1)$, $S(1,3)$, $S(2,3)$, and $S(3,3)$. At first, it is sufficient to calculate only $S(1,0)$ and $S(1,1)$,
\bea
S(1,0)&=&\frac{\coth(E_1/2T)}{2E_1},\\
S(1,1)&=&\frac{1}{2E_1E_2}\frac{E_1\coth(E_2/2T)-E_2\coth(E_1/2T)}{E_1^2-E_2^2},
\eea
since from these the rest can be obtained via differentiation:
\bea
S(n>1,0)&=&\frac{(-i)^{n-1}}{(n-1)!}\frac{\partial^{n-1}S(1,0)}{\partial(E_1^2)^{n-1}},\\
S(n>1,m>1)&=&\frac{(-i)^{n-1}}{(n-1)!}\frac{(-i)^{m-1}}{(m-1)!}\frac{\partial^{n-1}}{\partial(E_1^2)^{n-1}}\frac{\partial^{m-1}S(1,1)}{\partial(E_2^2)^{m-1}}.
\eea
Some terms in (\ref{Eq:flow_Ck}) contain sums in which the two energy values might be equal: $E_1=E_2$. These terms however are always multiplied by zero; therefore the appropriate limits are not listed here.
\end{widetext}


\begin{thebibliography}{9}
\bibitem{gellmann60} M. Gell-Mann and M. Levy, Nuovo Cimento {\bf 16}, 705 (1960).
\bibitem{chan73}  L.–H. Chan and R. W. Haymaker, Phys. Rev. D {\bf 7}, 402 (1973).
\bibitem{lenaghan00} J.T. Lenaghan, D.H. Rischke, and J. Schaffner-Bielich, Phys. Rev. D {\bf 62}, 085008 (2000).
\bibitem{roder03} D. R\"oder, J. Ruppert, and D.-H. Rischke, Phys. Rev. D {\bf 68}, 016003 (2003).
\bibitem{herpay06} T. Herpay and Zs. Sz\'ep, Phys. Rev. D {\bf 74}, 025008 (2006). 
\bibitem{kovacs07} P. Kov\'acs and Zs. Sz\'ep, Phys. Rev. D {\bf 75}, 025015 (2007).
\bibitem{kovacs08} P. Kov\'acs and Zs. Sz\'ep, Phys. Rev. D {\bf 77}, 065016 (2008).
\bibitem{schaefer09} B.-J. Schaefer and M. Wagner, Phys. Rev. D {\bf 79}, 014018 (2009).
\bibitem{jakovac10} A. Jakov\'ac and Zs. Sz\'ep, Phys. Rev. D {\bf 82}, 125038 (2010).
\bibitem{mitter13} M. Mitter and B.-J. Schaefer, Phys. Rev. D {\bf 89}, 054027 (2014). 
\bibitem{fukushima10} K. Fukushima and T. Hatsuda, Rep. Prog. Phys. {\bf 74}, 014001 (2011).
\bibitem{pisarski84} R. D. Pisarski and F. Wilczek, Phys. Rev. D {\bf 29}, 338 (1984).
\bibitem{berges02} J. Berges, N. Tetradis, and C.Wetterich, Phys. Rep. {\bf 363},
223 (2002).
\bibitem{fukushima10b} K. Fukushima, K. Kamikado, and B. Klein, Phys. Rev. D {\bf 83}, 116005 (2011).
\bibitem{kleinert06} H. Kleinert, Europhys. Lett. {\bf 74}, 889 (2006).
\bibitem{pelissetto13} A. Pelissetto and E. Vicari, Phys. Rev. D {\bf 88}, 105018 (2013).
\bibitem{nakayama14} Y. Nakayama and T. Ohtsuki, arXiv:1407.6195.
\bibitem{aoki12} S. Aoki, H. Fukaya, and Y. Taniguchi, Phys. Rev. D {\bf 86}, 114512 (2012).
\bibitem{aoki14} S. Aoki, H. Fukaya, and Y. Taniguchi, Proc. Sci. LATTICE2013 (2013) 139 [arXiv:1312.1417].
\bibitem{nakano09} E. Nakano, B.-J. Schaefer, B. Stokic, B. Friman, and K. Redlich, Phys. Lett. B {\bf 682}, 401 (2010).
\bibitem{fejos13} G. Fej\H{o}s, Phys. Rev. D {\bf 87}, 056006 (2013).
\bibitem{paterson81} A. J. Paterson, Nucl. Phys. {\bf B190}, 188 (1981).
\bibitem{wetterich93} C. Wetterich, Phys. Lett. {\bf B301}, 90 (1993).
\bibitem{pawlowski07} J.M. Pawlowski, Ann. Phys. {\bf 322}, 2831 (2007).
\bibitem{litim01} D. F. Litim, Phys. Rev. D {\bf 64}, 105007 (2001).
\bibitem{blaizot06} J.-P. Blaizot, A. Ipp, R. Mendez-Galain, and N. Wschebor, Nucl. Phys. {\bf A784}, 376 (2007).
\bibitem{blaizot10} J.-P. Blaizot, A. Ipp, and N. Wschebor, Nucl. Phys. {\bf A849}, 165 (2011).
\bibitem{patkos12} A. Patk\'os, Mod. Phys. Lett. A {\bf 27}, 1250212 (2012).
\bibitem{litim94} D. Litim, N. Tetradis, and C. Wetterich, Mod. Phys. Lett. A {\bf12}, 2287 (1997).
\bibitem{litim96} D. Litim, Phys. Lett. B{\bf 393}, 103 (1997).
\bibitem{ringwald90} A. Ringwald and C. Wetterich, Nucl. Phys. {\bf B334}, 506 (1990).
\end{thebibliography}
\end{document}